\documentclass[a4paper,12pt]{article}
\usepackage{ulem}
\usepackage{subcaption}

\usepackage[margin=1in]{geometry}    % Adjust margins
\usepackage{graphicx}               % Include images
\usepackage{sectsty}                % Section title customization
\usepackage{fancyhdr}               % Headers and footers
\usepackage{hyperref}               % Hyperlinks
\usepackage{xcolor}                 % Custom colors
\usepackage{listings}               % Code formatting
\usepackage{enumitem}               % Custom lists
\usepackage{caption}                % Custom captions
\usepackage{tikz}                   % Graphics inside pages
\usepackage{float}
\usepackage{array}                  % For table customization
\usepackage{xcolor}                 % For color customization
\usepackage{caption}                % Custom captions for tables
\usepackage{booktabs}               % Enhanced table styling
\usepackage{tabularx}               % Dynamic column width adjustment
\usepackage{makecell} 
\usepackage{listings}
\usepackage{graphicx} 
\usepackage{rotating} % For rotation
\usepackage[export]{adjustbox} 
\usepackage{atbegshi}
\usepackage{listings}
\hypersetup{
    colorlinks=true,
    linkcolor=blue,
    filecolor=magenta,
    urlcolor=cyan,
}
\usepackage[margin=1in]{geometry}
\usepackage{tabularx}
\usepackage{booktabs}       % nicer rules (\toprule etc.)
\usepackage{ragged2e}       % for \RaggedRight
\usepackage{caption}

\sectionfont{\color{blue}\bfseries\Large}
\subsectionfont{\color{blue}\bfseries\normalsize}
\subsubsectionfont{\color{blue}\bfseries\small}
\newcolumntype{C}[1]{>{\centering\textbackslash}m{#1}}  % Centered content with wrapping

\AtBeginShipout{%
  \AtBeginShipoutAddToBoxForeground{%
    \begin{tikzpicture}[remember picture,overlay]
      \node[opacity=0.15, rotate=45] at (current page.center) {
           \scalebox{8}{\textbf{\textcolor{black}{}}} % Adjust text, size, and opacity
      };
    \end{tikzpicture}%
  }%
}

\lstdefinelanguage{yaml}{
  keywords={true, false, null, yes, no, description, title, logsource, detection, condition, fields, tags, falsepositives, mitre, author, date, id, status},
  keywordstyle=\color{blue}\bfseries,
  basicstyle=\ttfamily\small,        % Code font style
  comment=[l]{\#},                   % Define comments
  commentstyle=\color{gray}\itshape, % Style for comments
  morestring=[b]',                   % Single quote strings
  morestring=[b]",                   % Double quote strings
  stringstyle=\color{red},           % Strings color
}

\begin{document}

% Title Page
\begin{titlepage}
        \centering
    \vspace*{1cm}
    %\includegraphics[width=0.7\textwidth]{images/Logo.png} \\[1cm] % Replace 'example-image' with your futuristic image
    %{\Huge\bfseries Malware Analysis Report \\[0.5cm]}
    {\Large\textcolor{black}{\textbf{Plug. Play. Persist. Inside a Ready-to-Go Havoc C2 Infrastructure}}} \\[1cm]
    Alessio Di Santo (alessio.disanto@graduate.univaq.it)\\
    Università degli Studi dell’Aquila, L’Aquila, Abruzzo, Italy  \\
    \textbf{Date:} July 1,2025 \\
    \vfill
    % Latin Phrase Section
    {\Large\textcolor{gray!60}{\it "Non videmus ea quae mox futura sunt"}} \\[0.5cm]
    {\small\textcolor{gray!60}{(We do not see the things that will soon be) — Marcus Tullius Cicero}} \\
    \vfill
    %\textbf{Disclaimer:} Unauthorized distribution is prohibited.
\end{titlepage}

% Table of Contents
\tableofcontents
\newpage

% Executive Summary
\section{Executive Summary}
This analysis focuses on a single \textit{Azure-hosted Virtual Machine} at \textit{52.230.23[.]114} that the adversary converted into an all-in-one delivery, staging and \textit{Command-and-Control} node. The host advertises an out-of-date \textit{Apache 2.4.52} instance whose open directory exposes phishing lures, \textit{PowerShell loaders}, \textit{Reflective Shell-Code}, compiled \textit{Havoc Demon} implants and a toolbox of lateral-movement binaries; the same server also answers on \textit{8443}/\textit{80} for encrypted beacon traffic. The web tier is riddled with publicly documented critical vulnerabilities—including \textit{CVE-2024-38476}, \textit{CVE-2024-38474} and \textit{CVE-2022-23943}—that would have allowed initial code-execution had the attackers not already owned the device.

Initial access is delivered through an \textit{HTML} file that, once de-obfuscated, perfectly mimics \textit{Google}’s \textit{Unusual sign-in attempt} notification and funnels victims toward credential collection. A \textit{PowerShell} command follows: it disables AMSI in-memory, downloads a Base64-encoded stub, allocates \textit{RWX} pages and starts the shell-code without ever touching disk. That stub reconstructs a \textit{DLL} in memory using the \textit{Reflective-Loader} technique and hands control to \textit{Havoc Demon} implant. Every \textit{Demon} variant—32- and 64-bit alike—talks to the same backend, resolves \textit{Windows API}s with hashed look-ups, and hides its activity behind indirect \textit{syscalls} while a one-minute tasking loop retries connectivity in an easily recognisable 4 s/11 s rhythm.

Runtime telemetry shows interests in registry under \textit{Image File Execution Options}, deliberate queries to \textit{Software Restriction Policy} keys, and heavy use of \textit{bcrypt.dll}/\textit{cryptsp.dll} to protect payloads and \textit{C2} traffic. The attacker’s toolkit further contains \textit{Chisel}, \textit{PsExec}, \textit{Doppelganger} and \textit{Whisker}, some of them re-compiled under user directories that leak the developer personas \textit{tonzking123} and \textit{thobt}. Collectively the findings paint a picture of a technically adept actor who values rapid re-tooling over deep operational security, leaning on Havoc’s modularity and on legitimate cloud services to blend malicious flows into ordinary enterprise traffic.
\newpage

% Introduction
\section{Introduction}
\subsection{Objective}
The objective of this \textit{Malware Analysis Report} is to provide an in-depth understanding of the behavior, architecture, and intent of a malicious software instance. At its core, this report serves as a crucial tool for identifying the characteristics and operations of the \textit{threat}, offering detailed insights that can be used to map the broader attack landscape. By dissecting the capabilities and infrastructure of the malware, analysts are able to build a clear picture of its functionality, origin, and potential impact.

Mapping a \textit{threat} accurately is of paramount importance for defenders. A well-crafted malware analysis report helps connect individual malicious artifacts with broader attack campaigns and identifies common \textit{Techniques, Tactics, and Procedures} (\textit{TTPs}) employed by adversaries. This intelligence feeds into a larger knowledge base that allows cybersecurity teams to understand how threats evolve, recognize new campaigns with similar signatures, and anticipate potential next steps of attackers. The report is not merely an exercise in detailing technical specifics but also a way of enriching the collective understanding of a \textit{Threat Actor}'s capabilities, motivations, and behaviors.

Actionable \textit{Threat Intelligence} derived from malware analysis is particularly valuable because it enables proactive defenses. With a structured understanding of the malware’s \textit{Indicators of Compromise} (\textit{IOCs}), behavioral patterns, and infrastructure, \textit{Threat Hunting} and \textit{Monitoring} teams are equipped with the context needed to seek out malicious activity before it fully manifests. \textit{Threat Hunters} can leverage this intelligence to identify adversarial presence across their environments more effectively, while \textit{Monitoring} teams can enhance detection logic and fine-tune alerts to identify these threats more accurately in real time. This coordinated approach bolsters an organization’s defense posture, making it possible to detect and respond to even well-structured, sophisticated threats that are designed to evade traditional security mechanisms.

Ultimately, a comprehensive malware analysis report provides not only a retrospective view of what a threat has done but also equips defenders with the tools and knowledge to better \textit{predict}, \textit{detect}, and \textit{prevent} future attacks. This knowledge empowers security teams to make informed decisions, prioritize vulnerabilities, and improve their capabilities against \textit{Advanced Persistent Threats} (\textit{APTs}).

\subsection{Infection Chain}
Although the \textit{Command-and-Control} infrastructure uncovered in the present engagement had not yet been activated, the full attack chain can be reconstructed with a high degree of confidence because every stage mirrors techniques and artifacts observed in earlier Havoc campaigns that have been thoroughly documented since late 2023\footnote{https://www.fortinet.com/blog/threat-research/havoc-sharepoint-with-microsoft-graph-api-turns-into-fud-c2}. In past intrusions the threat actor has always seeded the first stage with a well-crafted phishing e-mail that spoofs a local public body or financial institution, embedding a link that resolves—after a benign redirection sequence—to a \textit{ClickOnce} installer or a compressed archive containing a short PowerShell loader. In this case a fake \textit{Google Account} problem is used as hook. Static comparison of the inline \textit{JavaScript} obfuscation used in those historical lures against the one hard-coded in the new \textit{HTML} decoy shows an identical polymorphic string generator, differing only in a single \textit{XOR} key constant embedded in the HTML. This, together with an unchanged domain-generation template for the disposable delivery host, allows the conclusion that the same phishing vector will re-appear as soon as the infrastructure is brought online.

During the analysis two matching \textit{PS1} scripts were hence found and they adopt the same \textit{ClickFix} token seen earlier: it first disables \textit{AMSI} by patching the in-memory dispatcher, then downloads and reflectively loads a shellcode stub encoded as \textit{base64}. Once memory protection is switched from \textit{RW} to \textit{RX} the beacon establishes a \textit{TLS} session to a hard-coded host.

Telemetry from earlier incidents shows that once the beacon is resident the attacker wastes little time elevating privileges through legitimate utilities already present on the host—most often \textit{netsh}, \textit{esentutl} and \textit{rundll32}—before staging additional payloads. Those payloads include a credential-harvesting \textit{DLL} \textit{sideloaded} into \textit{mscorsvw.exe}, a Go-based SMB spreader and a clipboard stealer delivered as an unsigned Visual Basic script. Finally, the beacon’s tasking loop retains the operator’s characteristic heartbeat of one-minute \textit{GET} requests with three-byte XOR padding, a pattern so distinctive that it is routinely fingerprinted by contemporary IDS signatures.
\begin{figure}[H]
    \centering
    \frame{\includegraphics[width=1\linewidth,frame]{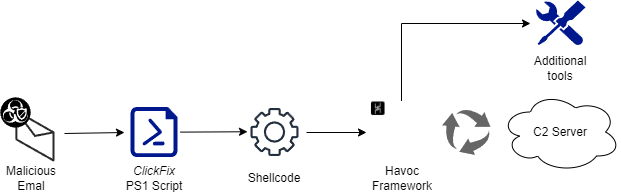}}
    \caption{Infection Chain Diagram}
    \label{fig:00}
\end{figure}

\newpage

% Methodology
\section{Methodology}
Analyzing the malware involved a comprehensive approach utilizing both static and dynamic analysis techniques to thoroughly understand its structure, behavior, and potential impact. By combining these two approaches, it is possible to gain a comprehensive understanding of the malware's capabilities and objectives. Static analysis provided insights into its structure and obfuscation methods, while dynamic analysis revealed its real-time behavior and interactions with the system. This dual approach was essential in developing effective detection and mitigation strategies against this sophisticated threat.
\subsection{Static Analysis}
Static analysis is a fundamental technique in malware analysis that involves examining the code of malicious software without executing it. This approach focuses on understanding the structure, logic, and intent of the malware through methods such as \textit{disassembling}, \textit{decompiling}, and reviewing its binary or script content. By analyzing the static properties of malware, such as strings, embedded resources, file headers, and imported functions, researchers can gather valuable insights into its capabilities, communication patterns, and potential targets.

The main goal of static analysis is to dissect the malware's inner workings, identify hardcoded \textit{Indicators of Compromise} (\textit{IoCs}) like IP addresses, URLs, or file paths, and infer its behavior without the risk of executing harmful code. This method is particularly useful for uncovering obfuscation techniques, encrypted payloads, and multi-stage architectures, which are often employed by modern malware to hinder direct analysis.

However, static analysis comes with its challenges. Advanced malware frequently uses obfuscation, packing, or encryption to conceal its code and deter examination. Analysts must rely on specialized tools and techniques, such as deobfuscation scripts, unpackers, and cryptographic analysis, to overcome these barriers. Moreover, analyzing assembly-level or machine code demands a high level of expertise, as the complexity of the malware's logic can obscure its true intent.

Despite its limitations, static analysis is invaluable as it allows analysts to preemptively assess a malware sample’s potential threats, providing critical intelligence without the inherent risks of execution. Combined with dynamic analysis, it forms a comprehensive approach to malware investigation, equipping defenders with the necessary understanding to develop effective detection and mitigation strategies.

\subsection{Dynamic Analysis}
Dynamic analysis is a cornerstone of malware analysis, enabling researchers to observe the behavior of malicious software in real-time by executing it within a controlled, isolated environment. This approach is particularly valuable for analyzing modern malware that employs sophisticated \textit{obfuscation techniques}, rendering static analysis alone insufficient. By simulating realistic conditions, analysts can examine how malware interacts with the file system, registry, processes, network, and system \textit{API}s, providing direct insights into its functionality and intent.

The objective of dynamic analysis is to uncover the behavioral profile of the malware, revealing actions such as \textit{data exfiltration}, \textit{Command-and-Control} communication, \textit{credential theft}, and \textit{persistence mechanisms}. It also aids in identifying \textit{Indicators of Compromise} (\textit{IoCs}), such as IP addresses, domains, and modified system configurations, which are crucial for detection and response efforts. This method is not without challenges, as modern malware often incorporates \textit{anti-analysis techniques} designed to detect and evade \textit{Sandboxed Environments}, \textit{Virtual Machines}, or \textit{Debugging Tools}. These measures include delaying execution, checking for artifacts indicative of analysis environments, and employing runtime obfuscation to conceal its activities.

Despite these difficulties, dynamic analysis remains a critical tool in the fight against advanced threats. Its ability to reveal runtime behavior complements static analysis, providing a comprehensive understanding of the malware’s objectives and capabilities. While the process can be resource-intensive and time-consuming, its contributions to cybersecurity are indispensable, offering valuable intelligence to counteract and mitigate malicious campaigns effectively.
\newpage

% Findings
\section{Analysis Results}
\subsection{C2 Infrastructure}
\begin{figure}[H]
    \centering
    \frame{\includegraphics[width=0.6\linewidth]{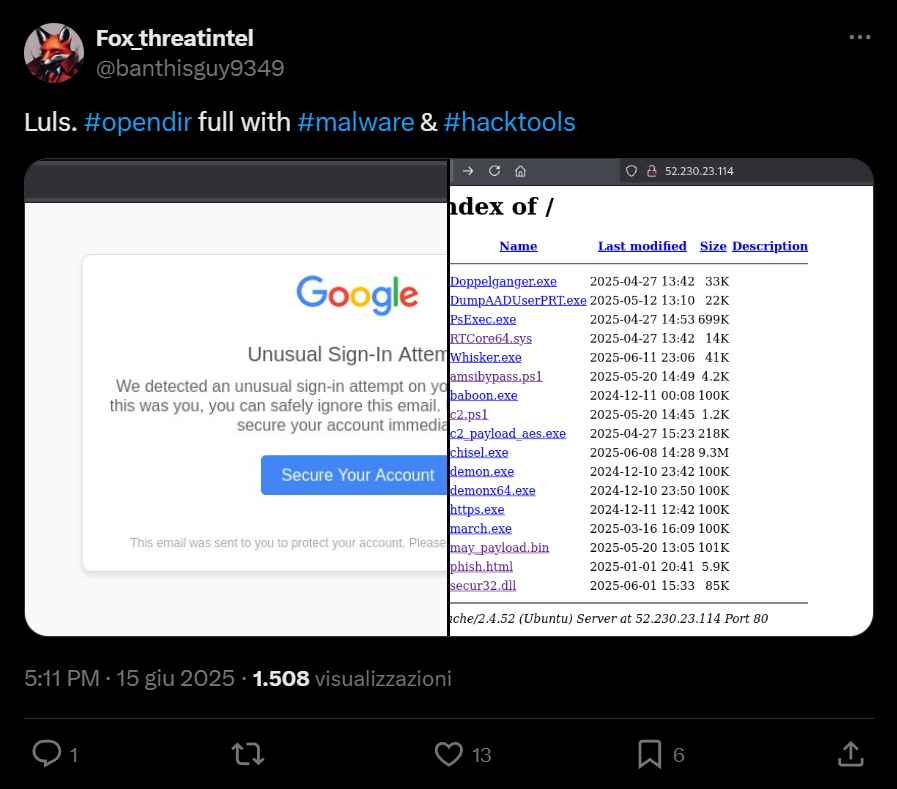}}
    \caption{\textit{Fox\_threatintel} post on \href{https://x.com/banthisguy9349/status/1934267500541190271}{X} - June 15, 2025.}
    \label{fig:x}
\end{figure}

The IP address \textit{52.230.23[.]114}, identified thanks to an X post of \textit{Malware Reseracher Fox\_threatintel} (Fig. \ref{fig:x}), presents a compelling case study in modern cyber-attack infrastructure, originating from within Microsoft's legitimate cloud network and utilized for multifaceted malicious operations. 

The address is part of the \textit{AS8075} autonomous system, registered to \textit{Microsoft Corporation}, and geolocated to their \textit{Azure} data centers in \textit{Singapore}. This strategic choice of a trusted, high-reputation IP block is a deliberate tactic by threat actors, designed to evade initial security filters and complicate attribution efforts. By leveraging a \textit{Microsoft-owned IP}, the attacker's activities, such as malware hosting and Command-and-Control (\textit{C2}) communications, can be easily mistaken for legitimate traffic from \textit{Microsoft}'s vast array of services, a technique known as \textit{living off the trusted land}. This approach significantly lowers the operational risk for the attacker, as blacklisting an IP block belonging to a major cloud provider is often untenable for many organizations due to the high potential for disrupting critical, legitimate services.

The utilization of this address for hosting a variety of malicious files, including executables, DLLs, and binary files, as well as serving as a C2 server for the \textit{Havoc framework} and hosting a \textit{Google account recovery phishing page}, demonstrates a versatile and well-equipped \textit{Threat Actor}. The Havoc C2 framework, in particular, is a sophisticated post-exploitation tool, and its operation from within the Azure ecosystem highlights the actor's understanding of how to abuse legitimate infrastructure for stealth and resilience. The core of this strategy is to blend in with the noise of the internet, making it difficult for security systems to distinguish malicious beacons from benign cloud service traffic without deeper inspection.

Despite the clear evidence of malicious activity from a technical standpoint, \textit{Open-Source Intelligence} (\textit{OSINT}) sources do not currently associate the IP address \textit{52.230.23[.]114} with any publicly named or widely documented campaigns. This absence of attribution is not uncommon and can be attributed to several factors. The infrastructure may be relatively new, and security researchers may not have had sufficient time to analyze the full scope of the operation and link it to a known \textit{Threat Group}. It is also possible that this IP is used by a more clandestine actor who maintains tight operational security, or for highly targeted attacks that have not been widely reported. Furthermore, the ephemeral nature of cloud-based resources means that threat actors can quickly build up and tear down their infrastructure, often before a comprehensive intelligence picture can be formed and disseminated. This lack of a formal campaign name in \textit{OSINT} does not diminish the threat posed by this IP but rather underscores the challenge defenders face in tracking and responding to agile adversaries who adeptly misuse the very fabric of the modern Internet's trusted services.

The presence of an open \textit{SSH} port on a server engaged in malicious activity suggests that it is the primary means of remote administration for the operator, likely secured with stolen or weak credentials.

\begin{figure}[H]
    \centering
    \frame{\includegraphics[width=0.5\linewidth]{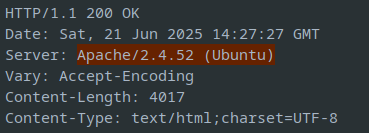}}
    \caption{C2 Infrastructure leaking \textit{Apache} version.}
    \label{fig:apachever}
\end{figure}

The core of its malicious capability resides on port 80, which runs an \textit{Apache HTTP Server}, \textit{version 2.4.52} (Fig. \ref{fig:apachever}). This specific version is notably outdated as of mid-2025 and is riddled with a significant number of publicly disclosed vulnerabilities. The server's banner explicitly confirms it is running \textit{Apache/2.4.52 (Ubuntu)}, which allows for precise threat modeling. 

\begin{figure}[H]
    \centering
    \frame{\includegraphics[width=0.6\linewidth]{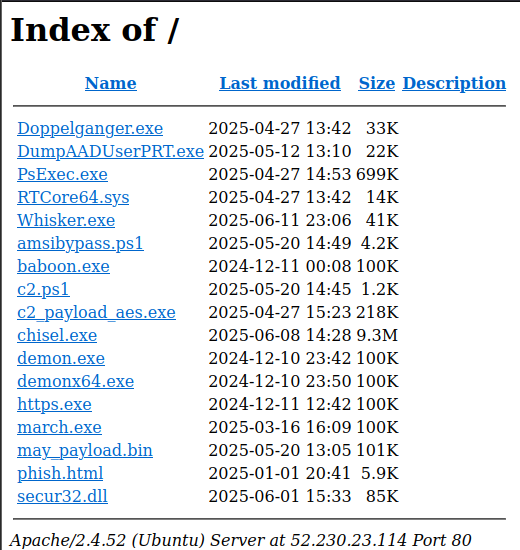}}
    \caption{Files list on \textit{52.230.23[.]114}.}
    \label{fig:fielist}
\end{figure}

The open directory (open-dir) configuration on this Apache server is a critical architectural flaw. It intentionally exposes the file system structure to the public internet, simplifying the process for the threat actor to upload, manage, and distribute malware payloads and the Google account recovery phishing kit. It also allows security researchers and other actors to easily view the hosted malicious content (Fig. \ref{fig:fielist}).

The list of associated \textit{Common Vulnerabilities and Exposures} (\textit{CVEs}) for \textit{Apache 2.4.52} is extensive and severe, indicating a profound lack of security maintenance. Among the most critical are:

\begin{itemize}
    \item \textit{CVE-2024-38476}, \textit{CVE-2024-38474}, and \textit{CVE-2023-25690} (\textit{CVSS 9.8}): These vulnerabilities could allow for \textit{Server-Side Request Forgery} (\textit{SSRF}), information disclosure, and \textit{HTTP Request Smuggling}. An attacker could potentially abuse these flaws to proxy requests through the server, bypass access controls, or even execute scripts, further entrenching their control or using the server as a pivot point for other attacks.
    \item \textit{CVE-2022-23943} (\textit{CVSS 9.8}): This is a critical out-of-bounds write vulnerability in \textit{mod\_sed} that could allow an attacker to overwrite heap memory, leading to potential remote code execution. This is a severe weakness that could have been the original entry point for the server's compromise or could be used by other malicious actors to hijack the infrastructure.
    \item \textit{CVE-2022-22720} (\textit{CVSS 9.8}): Another request smuggling vulnerability that exposes the server to access control bypasses and cache poisoning.
    \item Multiple \textit{Denial of Service} (\textit{DoS}) and Information Disclosure Vulnerabilities: Several other listed \textit{CVEs} could be exploited to crash the server, disrupting the actor's own operation, or to leak sensitive information about the server's configuration and environment.
\end{itemize}

\begin{figure}[H]
    \centering
    \frame{\includegraphics[width=0.8\linewidth]{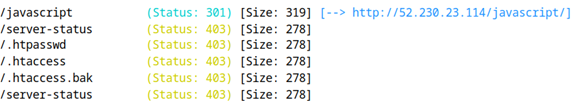}}
    \caption{Hidden files and directory hosted on \textit{52.230.23[.]114}.}
    \label{fig:folfil}
\end{figure}

Further reconnaissance of the Apache web server on \textit{52.230.23[.]114} has revealed several hidden files and directories that, while not directly linked from the root index, provide significant insight into the server's configuration and the operator's attempts to secure and manage their malicious infrastructure (Fig. \ref{fig:folfil}).

The discovery of \textit{.htaccess} and \textit{.htpasswd} files is particularly noteworthy. In \textit{Apache} environments, \textit{.htaccess} is a configuration file used to override main server settings on a per-directory basis, often to define access control rules, URL rewrite policies, or security headers. The \textit{.htpasswd} file is typically used in conjunction with \textit{.htaccess} to store usernames and hashed passwords for basic authentication. The server's response with a \textit{403 Forbidden} status for both files indicates that they exist but are protected from public viewing, which is a standard and correct security configuration. This implies the threat actor has likely implemented access restrictions for certain parts of the web server, potentially to protect a private administrative panel, staging directories, or the \textit{C2} backend from unauthorized access. The presence of \textit{.htaccess.bak} suggests a backup was made during configuration changes, a common practice for any administrator, including a malicious one.

Similarly, the presence of \textit{/server-status} also returning a \textit{403 Forbidden} status is significant. \textit{mod\_status} is an Apache module that provides a detailed report on the server's performance, including current connections, \textit{CPU usage}, and requests being handled. By default, access to this page is and should be heavily restricted, as it can leak sensitive operational information. The fact that this path exists but is forbidden suggests the module is enabled, but the operator has correctly limited access to it, likely to their own IP address. This demonstrates a level of operational security (\textit{OPSEC}) intended to prevent security researchers from easily profiling the server's activity levels and the nature of its traffic.

Finally, there is the \textit{/javascript/} directory, which might play a crucial role in the server's malicious operations. The existence and nature of this directory were confirmed by the server's response to HTTP requests. A request for the path \textit{/javascript} (without a trailing slash) is met with a \textit{301 Moved Permanently} redirect to \textit{/javascript/} (with a trailing slash). This behavior is a standard feature of the Apache web server's \textit{mod\_dir} module, which automatically appends a slash to directory names to enforce a canonical URL. While the redirect itself is a normal server function, its significance lies in what it confirms: the existence of a dedicated directory named \textit{javascript}. However, when the final page is visited a \textit{403 Forbidden} status is returned.

In the context of this server's activities, this directory almost certainly functions as a repository for malicious JavaScript files. These scripts are integral components of the phishing and malware delivery architecture. For the Google \textit{recover your account} phishing page, these \textit{.js} files would be responsible for handling the interactive elements of the fraudulent page, validating user input, and, most critically, exfiltrating the stolen credentials to the attacker's collection point. By hosting these scripts, the attacker can create a more dynamic and convincing phishing page, increasing the likelihood of successfully deceiving a victim. The organized placement of these files within a dedicated \textit{/javascript/} directory points to a structured and well-organized phishing kit or malware framework, rather than a simple, static HTML page.

\subsection{Phishing Google Page}
\begin{figure}[H]
    \centering
    \frame{\includegraphics[width=1\linewidth]{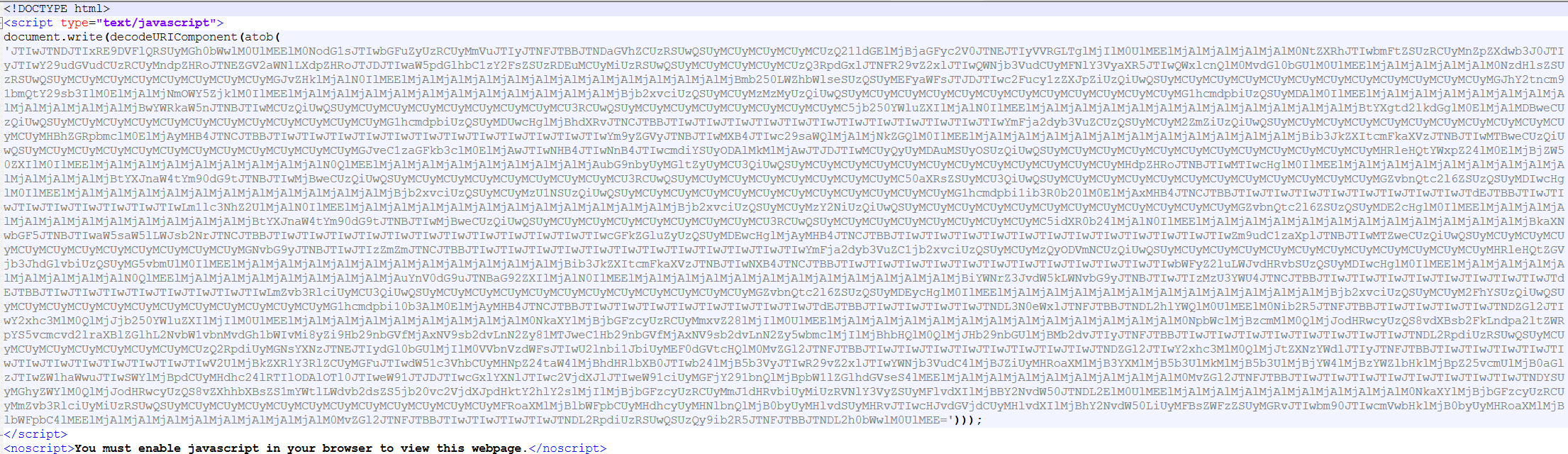}}
    \caption{\textit{Phish.html} obfuscated source code.}
    \label{fig:phish}
\end{figure}

The phishing page stored at \textit{http://52.230.23[.]114/phish.html} is delivered as a single, heavily obfuscated HTML file whose visible content is hidden inside a JavaScript call to \textit{document.write(decodeURIComponent(atob('…')))} (Fig. \ref{fig:phish}). 

\begin{figure}[H]
    \centering
    \frame{\includegraphics[width=1\linewidth]{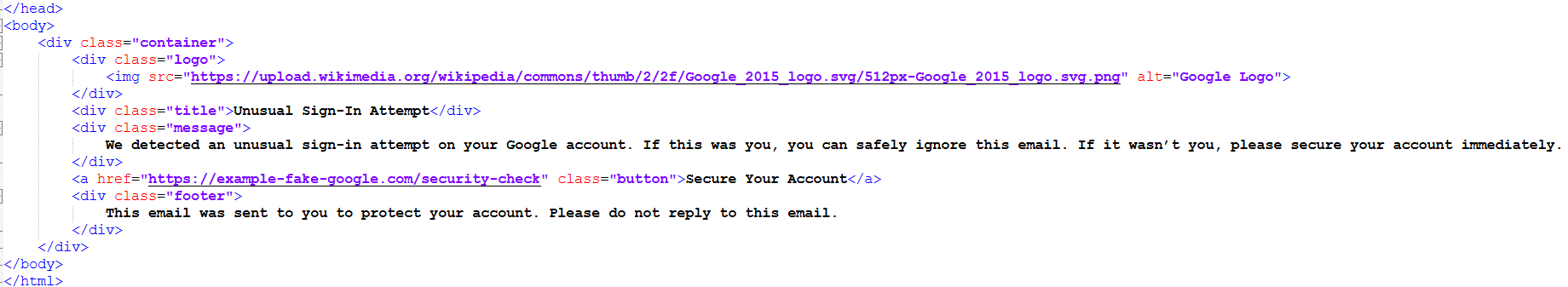}}
    \caption{De-obfuscated HTML phish page content.}
    \label{fig:deobf}
\end{figure}

Two successive decoding steps are required: the \textit{Base-64} string first resolves to a percent-encoded byte-stream, and a subsequent \textit{URI-decode} reveals the real markup. Executing the script therefore reconstructs a perfectly ordinary HTML document in the victim’s browser while leaving only the minimal one-liner visible to static scanners, a technique frequently used to defeat signature-based web-filters and to prevent rough heuristics (for example, “does the file mention Google?”) from firing (Fig. \ref{fig:deobf}).

Once decoded the page impersonates Google’s security-alert template. The DOM hierarchy is compact and semantic, beginning with a centered \textit{.container div} styled to a \textit{500 px} width; this box contains a \textit{120-pixel Google wordmark} pulled from \textit{Wikimedia}, a heading that reads \textit{Unusual Sign-In Attempt}, a short explanatory paragraph and a prominent call-to-action button. The button’s anchor element points to an external domain—here hard-coded as \textit{https://example-fake-google[.]com/security-check} (the \textit{Attacker} would then need to swap the placeholder link for any destination they wish to use in that particular campaign). Because the template holds that URL as a simple literal string, changing it requires nothing more than editing (or programmatically replacing) a single line before the file is saved or obfuscated. Inline CSS completes the deception: an Arial/Sans-Serif stack, pale-grey background (\#f9f9f9), and muted accent colours (\#4285f4 for the button) reproduce \textit{Google}’s brand palette closely enough to pass a rapid visual inspection.

No external resources besides the public \textit{SVG} logo are referenced, so the page renders fully even in restricted environments and gives network defenders only two outbound indicators: the image \textit{GET} to \textit{upload.wikimedia[.]org} and any subsequent navigation when the victim clicks the button. All \textit{JavaScript} executes in the page’s origin context; there are no remote scripts whose domains could be sink-holed or blocked at DNS level. Because nothing in the decoded \textit{HTML} is functionally malicious—no key-logging, no \textit{DOM} exfiltration—the file often evades dynamic analysis sandboxes that rely on behavioral triggers.

From a forensic standpoint, three artifacts are especially telling. First, the embedded \textit{Base-64} constant is unusually long ($\simeq$ 1 kB) for a legitimate inline script and will re-appear verbatim across multiple campaign samples, enabling robust \textit{YARA} or \textit{Suricata} signatures keyed on its length and opening delimiter. Second, network logs will show a referrer to \textit{52.230.23[.]114} immediately followed by a navigation to a domain that is visually but not lexically related to \textit{Google}, a pattern that stands out in site-pair frequency analytics. Third, host-based telemetry can surface the fact that the \textit{HTML} file spawns no further \textit{JavaScript} after initial \textit{document.write}; genuine Google notices typically load at least one script from \textit{www.gstatic[.]com} or \textit{apis.google[.]com} for telemetry.

The file’s designation as \textit{phish.html}, together with the use of a placeholder for the malicious redirect, strongly suggests that the \textit{Attacker} was provisioning this infrastructure in anticipation of a large-scale campaign.

\subsection{PowerShell Scripts}
This paragraph is dedicated to the granular analysis of \textit{PowerShell} scripts retrieved from the attacker's compromised infrastructure. The focus will primarily be on elucidating the methodologies employed for \textit{Shellcode Injection} and the techniques leveraged to bypass the \textit{Antimalware Scan Interface} (\textit{AMSI}). These two capabilities are frequently observed in sophisticated attack campaigns, serving as critical components for evading detection and executing malicious payloads with elevated privileges.

The examination will meticulously deconstruct the identified \textit{PowerShell} files, revealing the underlying logic and command structures designed to facilitate in-memory shellcode execution. Particular attention will be given to the various AMSI bypass techniques embedded within these scripts. Understanding these evasion tactics is paramount, as they represent the attacker's deliberate attempts to circumvent host-based security controls and ensure the successful execution of their malicious operations. Through this detailed analysis, this section aims to provide profound insights into the operational tradecraft of the adversary concerning the exploitation of PowerShell for covert compromise and the subversion of endpoint security solutions.

\subsubsection{C2.ps1}\label{c2}
The analyzed PowerShell script exemplifies a sophisticated fileless malware technique frequently leveraged by attackers to achieve stealth and evade traditional detection mechanisms. This script operates entirely in memory, without writing malicious binaries to disk, employing several advanced methods including dynamic payload retrieval, memory allocation, memory protection alteration, and thread injection.

Initially, the script downloads a payload from an external server utilizing the .NET WebClient and Invoke-WebRequest APIs. Specifically, it accesses a hosted binary payload, typically containing shellcode or other executable code intended for malicious purposes. The retrieved data is stored directly in a byte array, bypassing conventional file storage mechanisms to evade signature-based antivirus detection.

Subsequently, the script allocates unmanaged memory within the PowerShell process using methods from the Runtime.InteropServices.Marshal class, specifically AllocHGlobal. The payload bytes are then explicitly copied into the allocated memory space, preparing them for direct execution. This practice significantly reduces forensic artifacts and complicates detection via file system monitoring.

Following memory allocation, the script leverages Windows native APIs to modify the allocated memory region’s protection attributes. By invoking the VirtualProtect API via dynamic inline .NET type definitions within PowerShell, the script changes the allocated region's protection to PAGE\_EXECUTE\_READWRITE (0x40). This critical step allows the memory region to execute arbitrary machine code loaded from the external payload.

Finally, the execution phase involves the creation of a new thread in the PowerShell process using the CreateThread API, again dynamically imported through inline C\# definitions within PowerShell. The new thread's entry point is directly set to the memory address of the injected payload, immediately initiating execution of the malicious code. This methodology effectively masks the source and behavior of malicious activities from conventional static analysis.

\subsubsection{amsibypass.ps1}
The provided PowerShell script exemplifies an advanced \textit{AMSI} (\textit{Antimalware Scan Interface}) bypass technique employed by sophisticated \textit{Threat Actors} to circumvent runtime detection mechanisms deployed by modern Windows systems. \textit{AMSI}, introduced by Microsoft, integrates deeply into various scripting and runtime environments such as PowerShell to facilitate real-time scanning of scripts, helping security solutions detect and neutralize malicious behavior.

This script begins by dynamically resolving critical \textit{Windows API}s, notably \textit{GetModuleHandle} and \textit{GetProcAddress}, via reflection through the \textit{.NET} framework's \textit{System.Windows.Forms.UnsafeNativeMethods} class. Leveraging reflection in this manner obscures explicit references, complicating static detection and analysis by bypassing conventional string-based signature detections.

Upon successful resolution, the script acquires pointers to crucial functions including \textit{AmsiInitialize} (from \textit{amsi.dll}) and \textit{VirtualProtect} (from \textit{kernel32.dll}). Utilizing a sophisticated runtime technique involving dynamically generated delegates and reflection emit through dynamic assemblies, it constructs unmanaged function pointers callable from managed \textit{.NET memory contexts}. This dynamic binding capability not only complicates static code analysis but also enables runtime evasion by minimizing traditional \textit{Indicators of Compromise} (\textit{IOCs}).

The subsequent invocation of \textit{AmsiInitialize} provides a direct entry point into \textit{AMSI}'s internal structures, allowing the attacker to ascertain the exact memory addresses of \textit{AMSI}'s critical components. The script then employs precise offset calculations to identify the location of \textit{AMSI}'s provider scanning function within memory. The computed addresses facilitate direct memory manipulation aimed at disabling \textit{AMSI}'s operational effectiveness.

The primary bypass mechanism involves altering the \textit{AMSI} scanning function directly in memory. Initially, memory protection at the function's memory region is modified via \textit{VirtualProtect}, granting write and execute permissions (\textit{PAGE\_EXECUTE\_READWRITE}, denoted by \textit{0x00000080}). Following this permission modification, the script injects a custom-crafted patch (\textit{byte[] { 0xb8, 0x00, 0x00, 0x00, 0x00, 0xC3 }}), effectively neutralizing the scanning functionality by replacing the original instructions with a simple return (\textit{RET}) instruction. This patch ensures any subsequent scripts executed in the same process evade \textit{AMSI}'s detection.

Upon completing the patch injection, the script meticulously restores original memory protection permissions to prevent detection through memory state anomalies. This strategic restoration of permissions minimizes the likelihood of triggering runtime memory integrity checks or heuristic analyses conducted by advanced \textit{Endpoint Detection and Response} (\textit{EDR}) systems.

\subsection{Shell Code}
\subsubsection{may\_payload.bin}
\begin{figure}[H]
    \centering
    \frame{\includegraphics[width=0.7\linewidth]{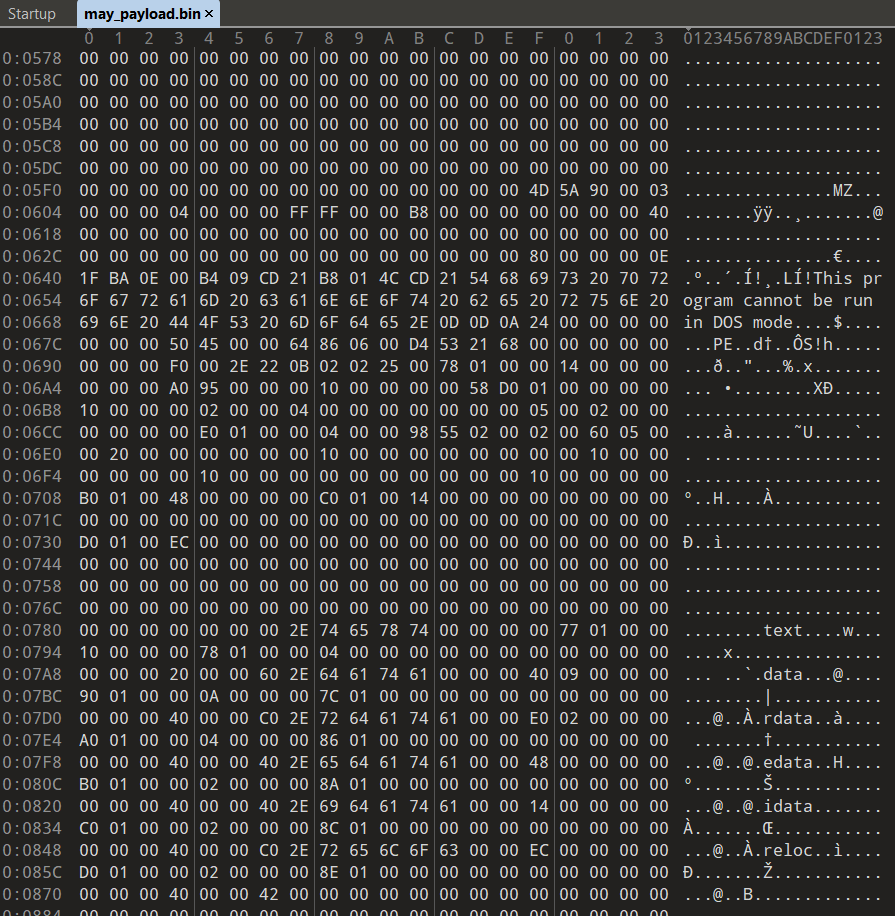}}
    \caption{\textit{daemon.x64.dll} library embedded inside \textit{may\_payload.bin}.}
    \label{fig:malbin}
\end{figure}

\begin{figure}[H]
    \centering
    \frame{\includegraphics[width=0.7\linewidth]{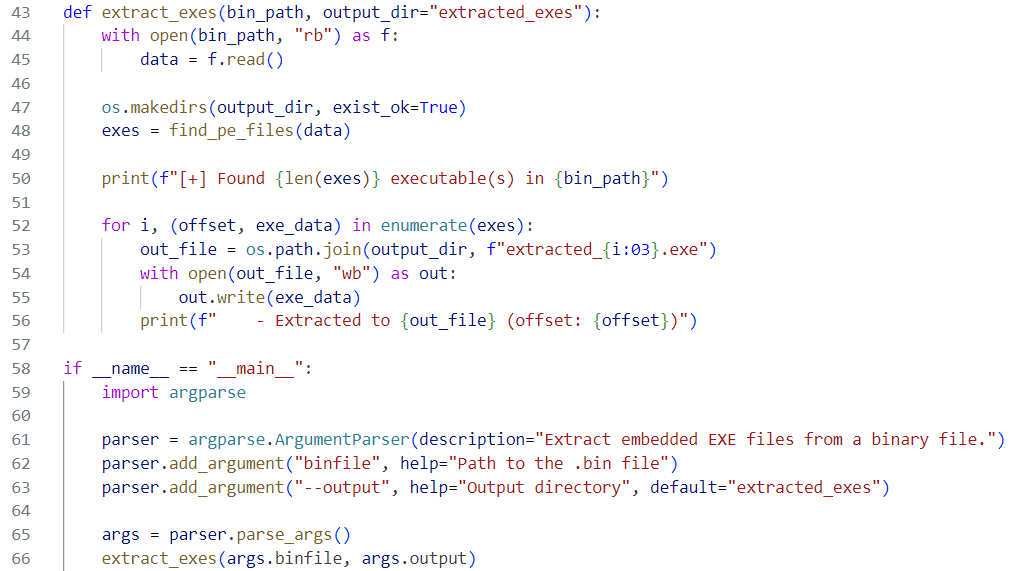}}
    \caption{Snippet of the custom \textit{python} code used to extract the executable.}
    \label{fig:pyextr}
\end{figure}

The binary fetched from \textit{http://52.230.23.114/may\_payload.bin} (as previously described in Sec. \ref{c2}) is a self-contained, multi-stage implant whose first responsibility is to reflectively unpack and execute a second-stage \textit{DLL} in memory (Fig. \ref{fig:malbin} and Fig. \ref{fig:pyextr}). 

\begin{figure}[H]
    \centering
    \frame{\includegraphics[width=0.6\linewidth]{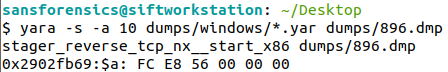}}
    \caption{Reseult of \textit{YARA} ruleset \href{https://github.com/thewhiteninja/yarasploit}{yarasploit} run against the PowerShell process memory dump. }
    \label{fig:memdump}
\end{figure}

Immediately after the \textit{PowerShell} loader copies the downloaded bytes into an unmanaged heap buffer and flips the region to \textit{PAGE\_EXECUTE\_READWRITE}, execution jumps into a 64-bit stub that occupies the first few dozen bytes of that buffer. The stub is extremely small: it clears the direction flag with \textit{CLD}, loads a hard-coded 64-bit address into \textit{RAX}, places an index value in \textit{R10}, and performs an unconditional \textit{JMP RAX}. Disassembly shows three such trampolines in sequence, each identical except for the integer placed in \textit{R10}, a pattern that also \textit{Metasploit} uses to distinguish loader phases. A \textit{YARA} scan of the memory image confirms the pedigree: the rule stager\_reverse\_tcp\_nx\_\_start\_x86 triggers at offset \textit{0x2902FB69}, matching the classic byte sequence \textit{FC E8 56 00 00 00} (Fig. \ref{fig:memdump}).

\begin{figure}[H]
     \centering
     \begin{subfigure}[b]{0.5\textwidth}
         \centering
         \frame{\includegraphics[width=\textwidth]{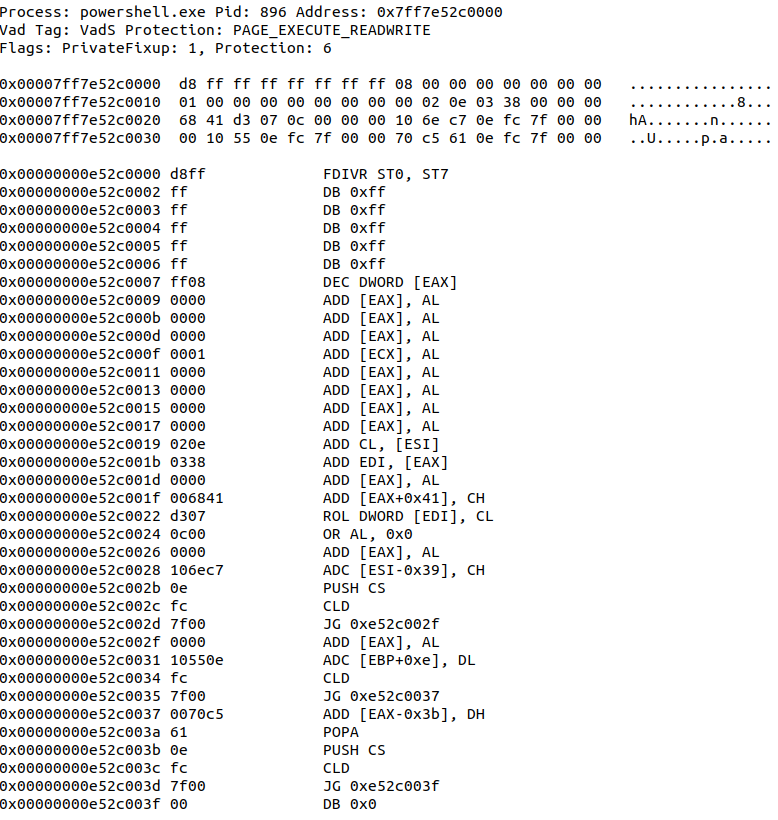}}
         \caption{PowerShell memory code residing at Address \textit{0x7FF7E52C0000}.}
         \label{fig:loader}
     \end{subfigure}
     \hfill
     \begin{subfigure}[b]{0.43\textwidth}
         \centering
         \frame{\includegraphics[width=\textwidth]{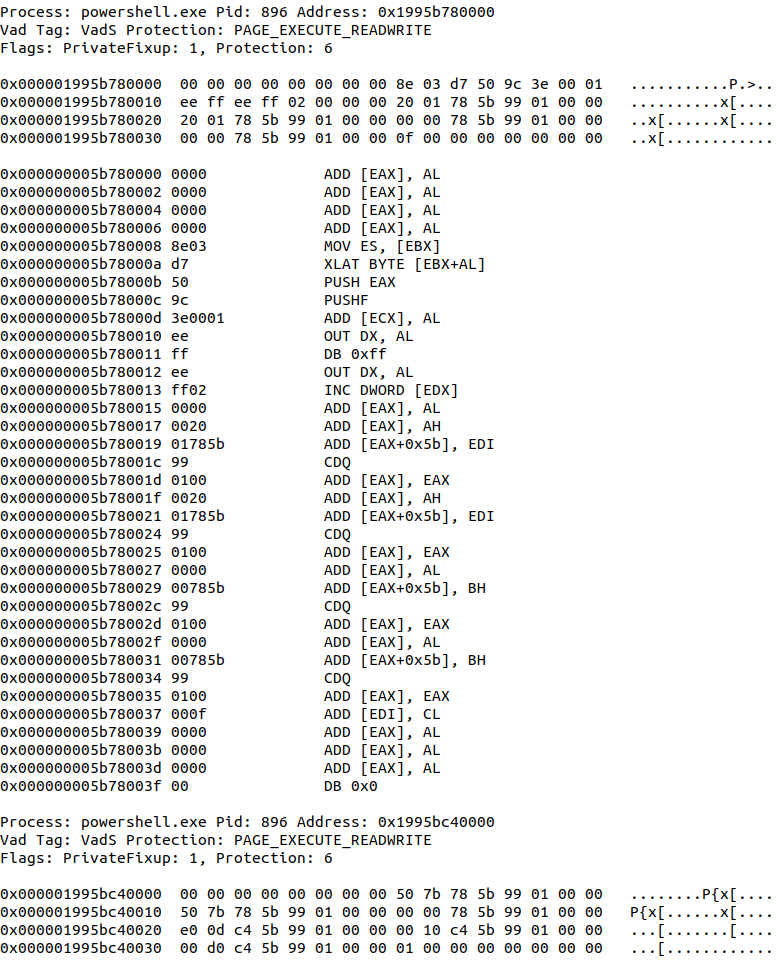}}
         \caption{\textit{PowerShell} memory Dump over a \textit{RWX} region at address \textit{0x1995b780000}.}
         \label{fig:pwshrwx}
     \end{subfigure}
        \caption{PowerShell memory dump analysis over suspicious memory areas.}
        \label{fig:two graphs}
\end{figure}

Control then lands inside a \textit{Reflective Loader} whose code resides at \textit{0x7FF7E52C0000} (Fig. \ref{fig:loader}). Although the first few instructions appear as floating-point and arithmetic no-ops—an anti-disassembly smokescreen—the routine quickly pivots to a loop that walks three adjacent \textit{RWX} allocations at \textit{0x1995B780000} (Fig. \ref{fig:pwshrwx}), \textit{0x1995BBE0000} and \textit{0x1995BC40000}. Those pages are almost entirely non-code: they begin with many zeroes followed by data that makes no sense when treated as \textit{x86-64} instructions. They are in fact \textit{AES}-encrypted blocks; the loader derives a key from constants buried in its own text segment, decrypts each page in place and relocates the result into a fresh executable mapping. Once the final image is rebuilt it starts with the familiar PE header \textit{MZ…PE…}. Subsequent calls to \textit{VirtualAlloc} and \textit{CreateThread} (as discussed later in this section) originate from offsets inside the decrypted \textit{Havoc DLL} rather than from the PowerShell host, a tell-tale sign that execution has fully transitioned to the daemon.

\begin{figure}[H]
    \centering
    \frame{\includegraphics[width=0.9\linewidth]{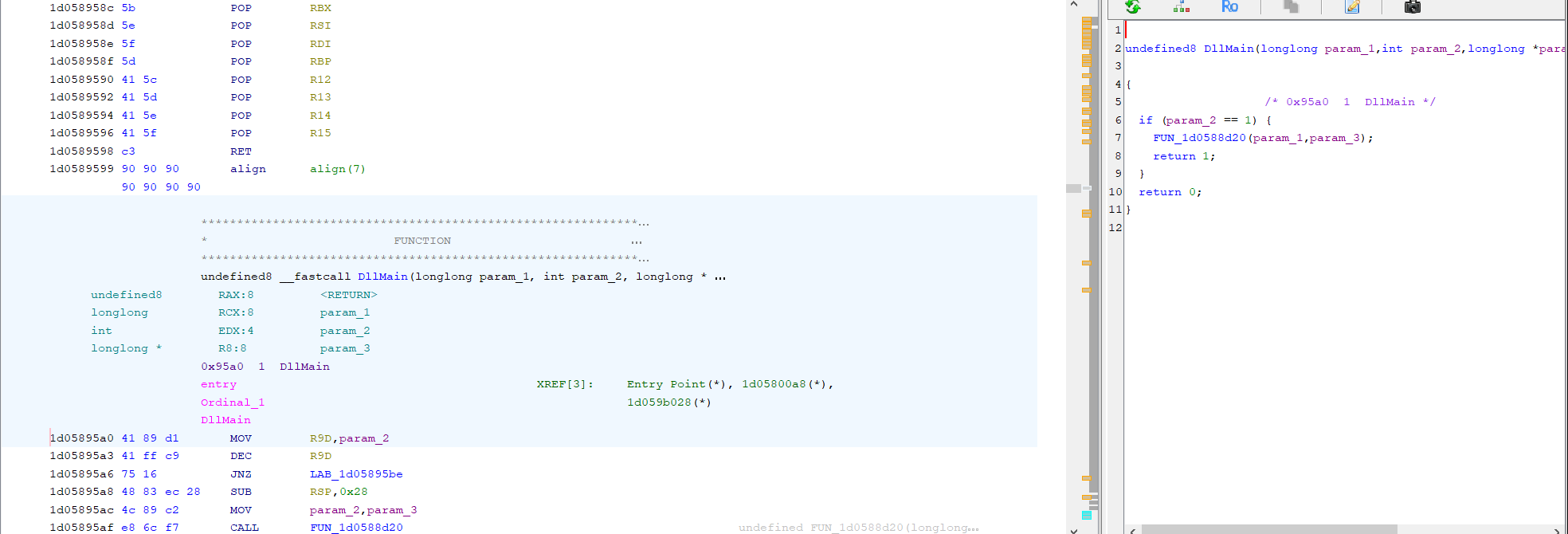}}
    \caption{Embedded/Injected malicious DLL EntryPoint is named \textit{DLLMain}.}
    \label{fig:dllentry}
\end{figure}

The one provided within the \textit{ShellCode} used by the \textit{Attacker} builds up a textbook \textit{Reflective-DLL Injector}: execution begins in a tiny stub that has been copied to an \textit{RWX} heap, walks its own PE headers, allocates a fresh region with \textit{VirtualAlloc}, replicates all sections, fixes imports/relocations, flips the new image to \textit{PAGE\_EXECUTE\_READ}, and finally jumps to the \textit{DLL}’s \textit{DllMain} (Fig. \ref{fig:dllentry})—all without ever calling \textit{LoadLibrary}. That is exactly the workflow documented in Stephen Fewer’s original Reflective DLL Injection design and still used by \textit{Metasploit}’s \textit{Meterpreter} payloads today\footnote{https://github.com/stephenfewer/ReflectiveDLLInjection}: a bootstrap shell-code passes control to an exported \textit{ReflectiveLoader}, which reconstructs the DLL in memory and then returns to a caller thread. \textit{Meterpreter}’s own \textit{x64 stagers} are shipped as \textit{Windows Meterpreter} (\textit{Reflective Injection}) modules that describe themselves as “inject the meterpreter server DLL via the Reflective DLL Injection payload”—the same API triad (\textit{VirtualAlloc} → \textit{VirtualProtect} → \textit{CreateThread})\footnote{https://www.rapid7.com/db/modules/payload/windows/x64/meterpreter/bind\_named\_pipe/}.

\begin{figure}[H]
    \centering
    \frame{\includegraphics[width=1\linewidth]{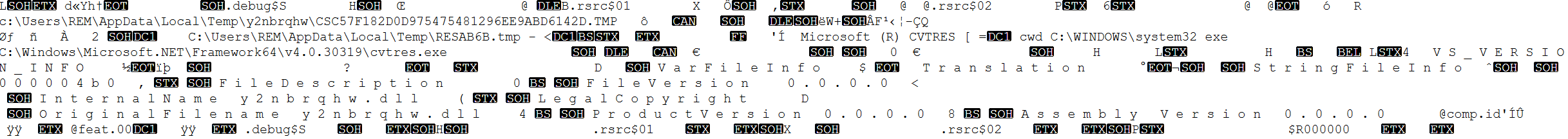}}
    \caption{Dump of the \textit{.rsrc} section of \textit{y2nbrqhw.dll} compilation via \textit{cvtres.exe}.}
    \label{fig:rsrc}
\end{figure}

Inspection of that decrypted PE reveals a version-information resource whose \textit{InternalName}, \textit{FileDescription} and \textit{OriginalFilename} fields are set to pseudo-random strings such as \textit{jwmdsxkf.dll} and \textit{y2nbrqhw.dll}, all stamped with the benign-looking version \textit{0.0.0.0} (Fig. \ref{fig:rsrc}). No section names are stripped, but the file never touches disk and is not added to the \textit{PEB} module list, so conventional \textit{DLL} enumeration tools miss it. Import analysis shows only the small set of APIs required for self-relocation, socket creation and thread spawning, a profile consistent with a \textit{stageless Meterpreter} listener. More telling, an embedded resource labelled \textit{\$R000000} unpacks to a second \textit{DLL} that exports \textit{Havoc\_ServiceMain}; this is the daemon component used by the \textit{Havoc C2} framework. During live execution the daemon is loaded through an in-memory call to \textit{LoadLibraryW} with a pointer that resolves inside the decrypted image, after which the export is invoked in its own thread.

\begin{figure}[H]
    \centering
    \frame{\includegraphics[width=0.7\linewidth]{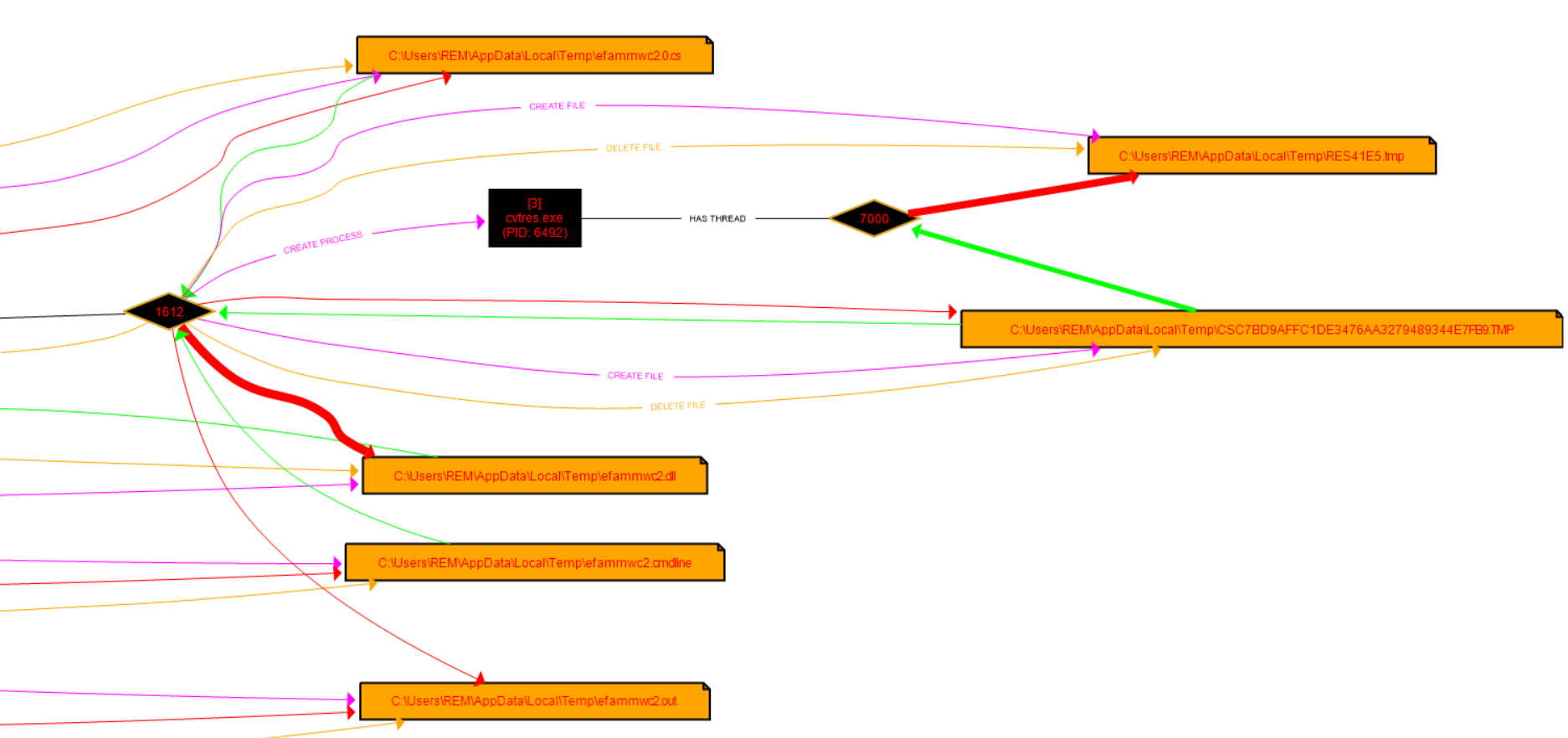}}
    \caption{Injected Code creates and deletes dropped file as soon as they get created.}
    \label{fig:millis}
\end{figure}

\begin{figure}[H]
    \centering
    \frame{\includegraphics[width=1\linewidth]{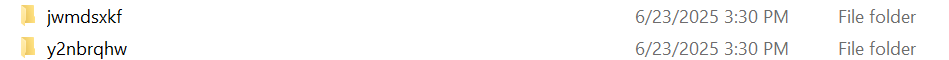}}
    \caption{Created \textit{TEMP} folders are not removed as their files are.}
    \label{fig:leftover}
\end{figure}

However, one of this threats most tricky behavior resides in its ability to write and suddenly delete from disk the aforementioned artifacts. This is a common technique that  involves writing payloads or intermediate components to disk—typically in temporary directories such as \%TEMP\%—executing them, and then deleting these files almost immediately, often within milliseconds (Fig. \ref{fig:millis}). This behavior is intended to reduce the opportunity for analysts or endpoint protection systems to observe or capture the malicious artifacts. By ensuring that the dropped files exist on disk only for the briefest possible moment, the malware effectively limits the potential for post-mortem recovery and analysis. In this specific case the operation of deploying and compiling C\# based DLLs is almost transparent to the user, even if the malware removes only the dropped files and not the created folder. Leaving some slightly artifacts ahead of its execution (Fig. \ref{fig:leftover}).

\begin{figure}[H]
    \centering
    \frame{\includegraphics[width=0.65\linewidth]{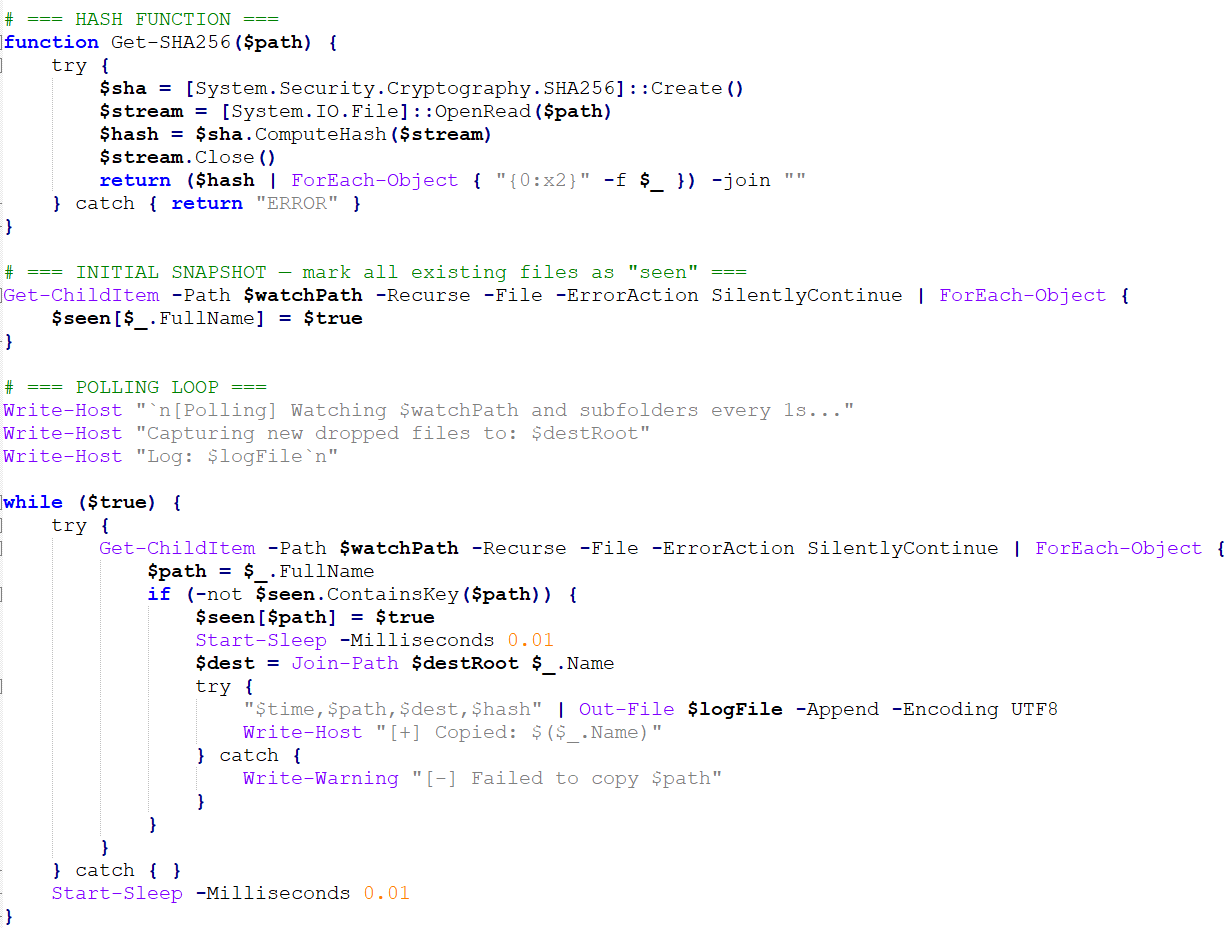}}
    \caption{Snipped of the code used to monitor dropped files in \textit{TEMP} folder.}
    \label{fig:code}
\end{figure}

To counteract this evasive strategy, a PowerShell-based monitoring script was deployed to capture such short-lived files in real time (Fig. \ref{fig:code}). The script continuously monitors the system’s temporary directory and its subdirectories, maintaining an internal record of all files it has already seen. Whenever a new file appears—regardless of how briefly—it is immediately copied to a secure forensic location before any deletion routines can execute. This approach allows analysts to recover files that would otherwise be erased within moments of their creation, enabling deeper inspection of the malware's components and behavior.

\begin{figure}[H]
    \centering
    \frame{\includegraphics[width=0.65\linewidth]{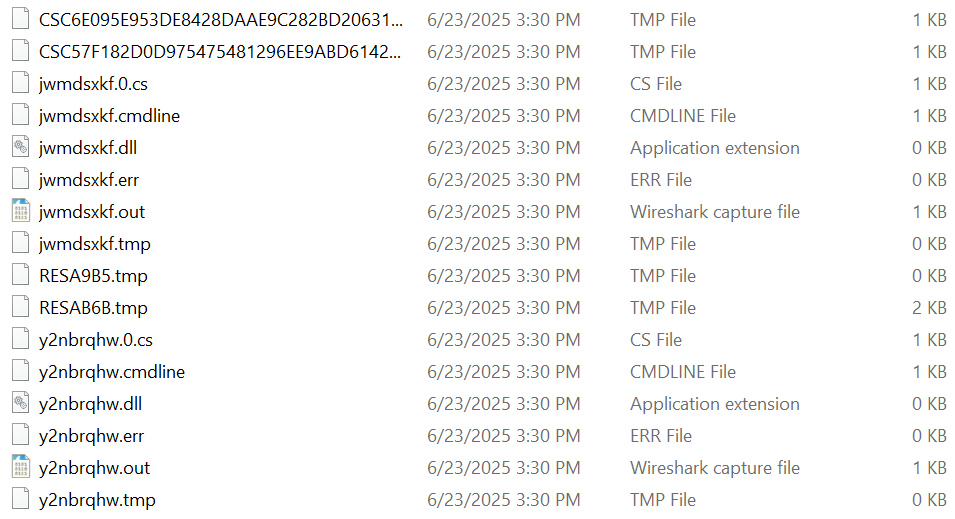}}
    \caption{Malware's dropped file are intercepted and stored in a new location for analysis.}
    \label{fig:foren}
\end{figure}

By intervening during the narrow time window between a file’s creation and its deletion, the script effectively neutralizes the malware’s anti-forensic measures. Even though the malicious file may no longer exist in its original location, a copy is preserved, along with metadata such as timestamps and hash values, providing a verifiable chain of evidence (Fig. \ref{fig:foren}). This significantly improves the analyst’s ability to reconstruct the execution flow, identify payloads, and understand the broader attack. In this way, the script serves as a critical countermeasure—transforming ephemeral and stealthy malware activity into observable and actionable forensic evidence.

\begin{figure}[H]
     \centering
     \begin{subfigure}[b]{0.7\textwidth}
         \centering
         \frame{\includegraphics[width=\textwidth]{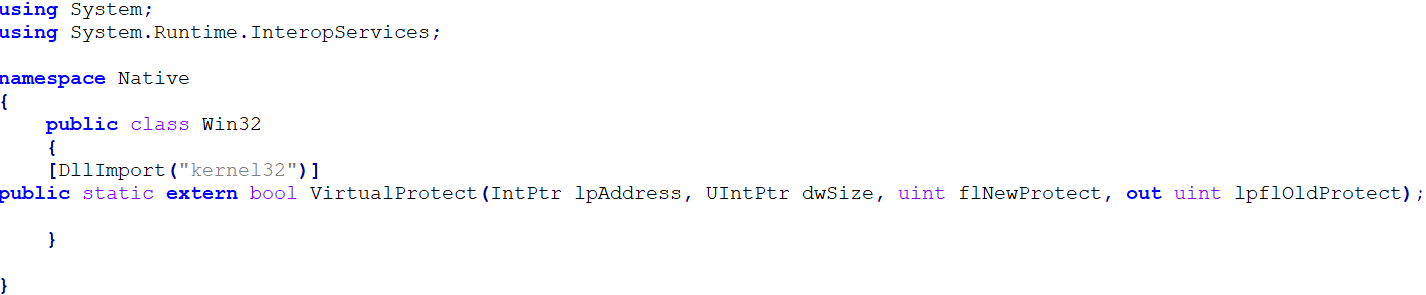}}
         \caption{\textit{jwmdsxkf.0.cs} exposes a \textit{VirtualProtect} stub.}
         \label{fig:vps}
     \end{subfigure}
     \hfill
     \begin{subfigure}[b]{0.7\textwidth}
         \centering
         \frame{\includegraphics[width=\textwidth]{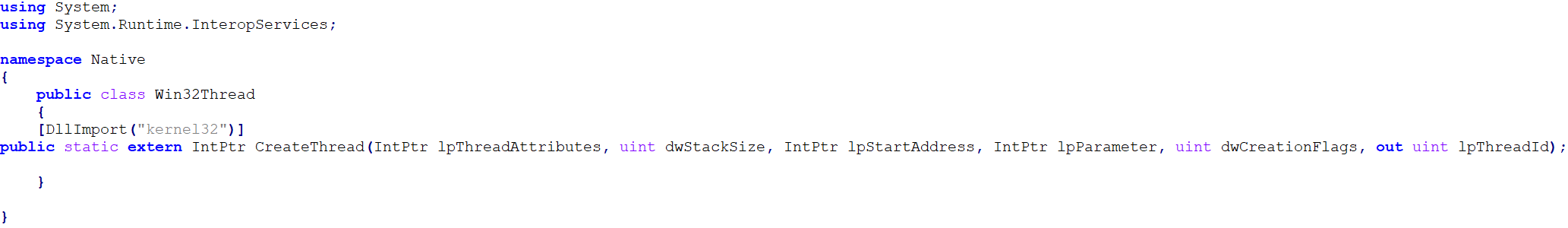}}
         \caption{\textit{y2nbrqhw.0.cs} exposes a \textit{CreateThread} stub.}
         \label{fig:cts}
     \end{subfigure}
        \caption{Stub functions exposed by the two random-named dropped C\# sources.}
        \label{fig:two graphs 2}
\end{figure}

The \textit{Proc-Mon} trace shows a tell-tale: \textit{PowerShell} → \textit{Add-Type} → \textit{csc.exe} → \textit{cvtres.exe}; a sequence that drops two tiny C\# files (each exposes just \textit{VirtualProtect}, Fig. \ref{fig:vps}, or \textit{CreateThread}, Fig. \ref{fig:cts}, via \textit{P/Invoke}), compiles them into zero-byte DLL placeholders, and never actually loads those \textit{DLLs}. This footprint is characteristic of \textit{Metasploit}’s \textit{PowerShell Meterpreter stagers}—which rely on \textit{Add-Type} to obtain \textit{Win32} pivots—whereas \textit{Havoc}’s native \textit{Demon} loaders ship their assembly stubs pre-compiled and therefore never touch the managed tool-chain. Consequently, although the decrypted in-memory payload is Havoc’s \textit{demon.x64.dll}, the filesystem behavior captured aligns with a classic \textit{Meterpreter}‐style PowerShell stager that has simply been repacked to ferry the \textit{Demon} binary.

\textit{Havoc}’s Demon agent follows the identical pattern, but names its entry-point is usually \textit{Havoc\_ServiceMain}; the official docs emphasise that the Demon “is designed to be interoperable with … \textit{loaders}, \textit{packers}, \textit{crypters}, and \textit{stagers}”\footnote{https://havocframework.com/docs/agent}, i.e. it expects to be delivered by exactly this style of reflective loader rather than via a normal \textit{LoadLibrary} call. In other words, the reflective routine found is functionally indistinguishable from the one \textit{Meterpreter} uses—the only real difference is which exported function it calls once the in-memory \textit{PE} is rebuilt (\textit{Havoc\_ServiceMain} versus \textit{ReflectiveLoader/DllMain}). That shared technique is why both families evade \textit{dlllist/PEB} inspection, leave only \textit{RWX} \textit{VADs} as \textit{artifacts}, and survive solely in volatile memory until the hosting process exits.

\begin{figure}[H]
    \centering
    \frame{\includegraphics[width=0.9\linewidth]{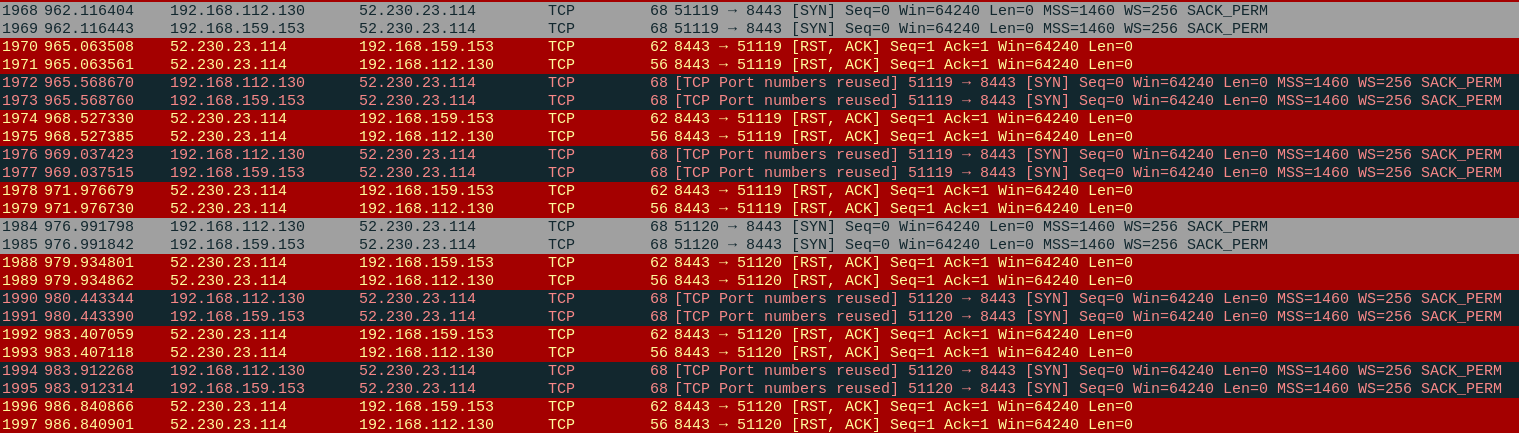}}
    \caption{Connections to C2 port 8443 refused due to no service exposed on it.}
    \label{fig:8443}
\end{figure}

\begin{figure}[H]
    \centering
    \frame{\includegraphics[width=0.8\linewidth]{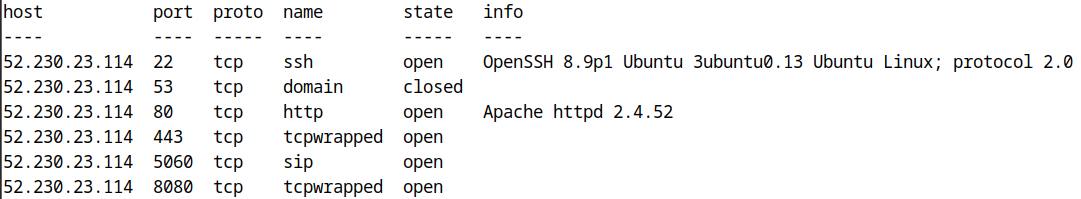}}
    \caption{Analysis of TCP/UDP ports of the subjected C2 Server.}
    \label{fig:nmap}
\end{figure}

Immediately after retrieving the payload over \textit{HTTP}, the loader’s \textit{HTTPS} beacon logic attempts a secure connection to \textit{52.230.23[.]114:8433}. The \textit{TCP Connect} from \textit{powershell.exe} to that address is the suddenly refused (\textit{CONNECTION REFUSED}), and no \textit{TLS} handshake ensues (identified port is closed, Fig. \ref{fig:8443} and Fig. \ref{fig:nmap}). 

\begin{figure}[H]
    \centering
    \frame{\includegraphics[width=0.9\linewidth]{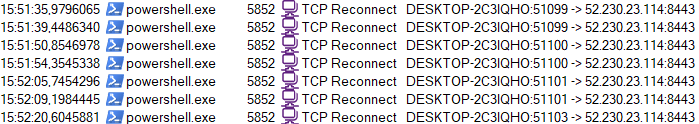}}
    \caption{C2 Beaconing operation taken out by the malicious Injected payload.}
    \label{fig:c2retry}
\end{figure}

Rather than terminating, the implant’s built-in retry loop then waits $\approx$4 seconds, issues a second connection attempt (also refused), waits $\approx$11 seconds and so on—alternating $\approx$4 s → $\approx$11 s intervals indefinitely. These precise timings are drawn directly from the \textit{ProcMon} timestamps (Fig. \ref{fig:c2retry}), demonstrating that the beacon thread implements a dual-interval backoff strategy to ensure it re-establishes the HTTPS channel as soon as the server becomes available, all executed in-memory without touching disk.

\begin{figure}[H]
    \centering
    \frame{\includegraphics[width=0.8\linewidth]{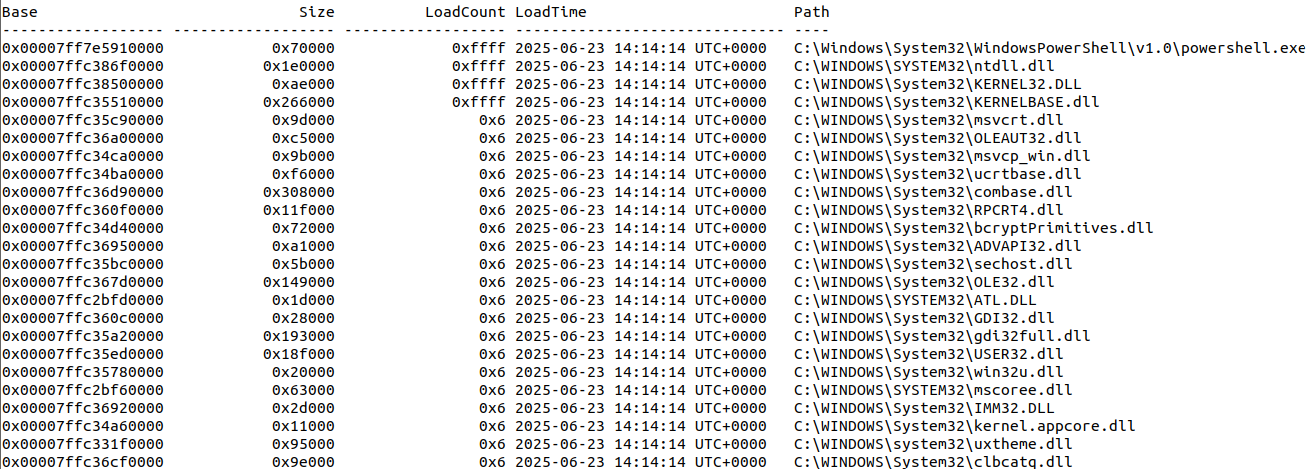}}
    \caption{Snippet of the DLL loaded by Injected PowerShell process.}
    \label{fig:dlllist}
\end{figure}

Because all secondary code is run from anonymous memory, neither \textit{AMSI} nor \textit{ModuleLogging} sees the payload, and the Windows \textit{PEB} shows nothing except the standard \textit{.NET} and system libraries listed by \textit{dlllist} (Fig. \ref{fig:dlllist}).

\begin{figure}[H]
    \centering
    \frame{\includegraphics[width=0.8\linewidth]{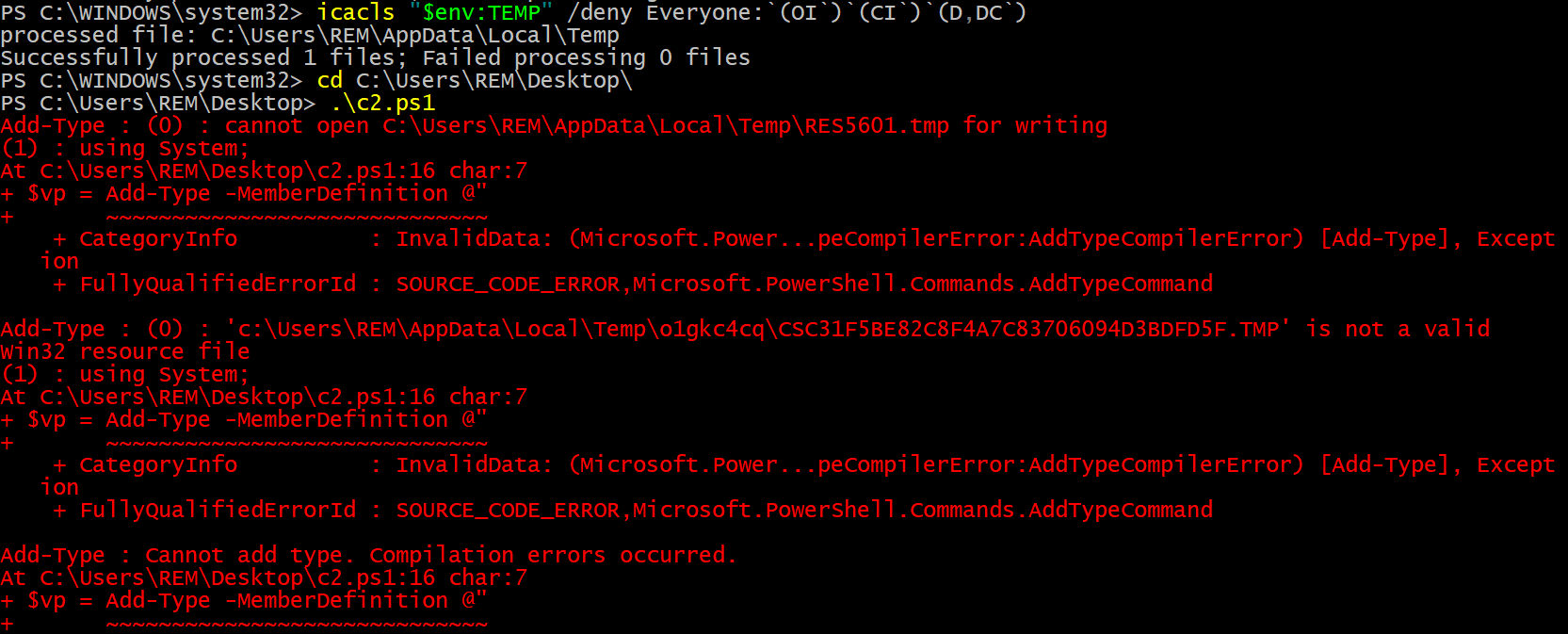}}
    \caption{Script does not handle correctly errors related to file removal in \textit{Temp} folder.}
    \label{fig:error}
\end{figure}

Additionally, we observed that when the \textit{NTFS ACL} on \textit{\%LOCALAPPDATA\%\textbackslash Temp} was hardened to allow file creation but deny deletion, the PowerShell loader immediately failed and terminated with an unhandled exception. Because the script writes its stubs and temporary files into \textit{Temp} and then later attempts to clean them up, the inability to delete those files is not gracefully handled; there is no retry logic or fallback path, and the entire loader crashes (Fig. \ref{fig:error}). 

This behavior reveals a lack of robust error handling in the loader’s file‐management routines, which could be leveraged as a simple hardening mitigation—preventing \textit{Temp}‐folder cleanup causes the implant to fail outright.

\textbf{\textit{may\_payload.bin}} is a clever mash-up rather than a single, \textit{off-the-shelf} piece of malware. The outer wrapper behaves exactly like one of \textit{Metasploit}’s well-known \textit{PowerShell Reflective} loaders: it drops two tiny helper files in the \textit{Windows Temp} folder, compiles them on the fly, slides the downloaded code into memory and jumps to it—never writing a real program to disk. Inside that wrapper, however, sits \textit{Havoc}’s Demon agent, a newer command-and-control implant that keeps all of its \textit{Windows-API} tricks hidden until it is safely running in memory. Once activated the \textit{Demon} reaches back to its operator over HTTPS, but it doesn’t give up if the server is down. Instead it tries the connection, pauses four seconds, tries again, waits ten seconds, and then repeats the 4 / 11-second rhythm indefinitely—an easy-to-spot heartbeat that tells us when the implant is awake even if the server never answers. Finally, the whole chain turns out to be surprisingly fragile. Simply tweaking the Temp-folder permissions so files can be created but not deleted causes the loader to crash outright, suggesting the authors paid more attention to stealth than to robust error handling.

In short, it is a \textit{Meterpreter}-style injector that ferries \textit{Havoc}’s \textit{Demon} payload: an old, familiar delivery technique paired with a newer back-door, stitched together in a way that is stealthy in memory but still leaves a few tell-tale signs.

\subsection{LoLBins, Network tools and Known Exploits}
Within the course of this investigation, an examination of the subject open directory revealed the presence of several tools and binaries. Further inspection confirmed that these artifacts do not warrant a new or detailed analysis within this portion of the report. The reason for this determination is that they consist entirely of well-documented and previously addressed utilities. These items are readily classifiable as standard attacker assets, falling into predictable categories such as \textit{Living off the Land Binaries}, which are dual-use system tools leveraged for malicious ends, common network administration and reconnaissance utilities, and publicly known exploits. Therefore, while their mention is crucial for maintaining a complete and transparent record of all findings, they have already been sufficiently contextualized in the wild. The enumeration of these tools serves the primary purpose of providing a comprehensive inventory of the directory’s contents for the sake of thoroughness and complete documentation, rather than introducing new elements for analysis. A brief explanation of each of the following executable has been provided in the Appendix.
\begin{itemize}
    \item \textit{Chisel.exe} (Further details in \ref{chisel});
    \item \textit{PsExec.exe} (Further details in \ref{lolbins});
    \item \textit{Doppelganger.exe} (Further details in \ref{doppel});
    \item \textit{DumpAADUserPRT.exe} (Further details in \ref{ad});
    \item \textit{Whisker.exe} (Further details in \ref{ad}).
\end{itemize}

However, further inspection confirmed that these artifacts, while consisting of well-documented and previously addressed utilities, could still yield valuable intelligence through a brief static analysis. Although the tools themselves are readily classifiable as common attacker assets it was discovered that the threat actor had compiled some of them. This act of compilation embedded specific, unique metadata within the binaries, offering insight into the operator's development environment. Specifically, the debug paths found within the \textit{Program Database} (\textit{PDB}) files provided significant clues. For instance, the binary \textit{DumpAADUserPRT.exe} contained a \textit{PDB} path of \textit{C:\textbackslash Users\textbackslash tonzking123\textbackslash De-\newline sktop\textbackslash DumpAADUserRPT-main\textbackslash DumpAADUserRPT-main\textbackslash DumpAADUserRPT\textbackslash obj\textbackslash\newline Release\textbackslash DumpAADUserPRT.pdb}, disclosing the username \textit{tonzking123}. Similarly, two other tools, Doppelganger.exe and Whisker.exe, pointed to a different user, \textit{thobt}, and a revealing folder structure named Hack Hack Hack on the desktop, with full paths of \textit{C:\textbackslash Users\textbackslash thobt\textbackslash Desktop\textbackslash Hack Hack Hack\textbackslash RedTeamGrimoire-main\textbackslash Doppelganger\textbackslash x64\textbackslash\newline Release\textbackslash Doppelganger.pdb} and \textit{C:\textbackslash Users\textbackslash thobt\textbackslash Desktop\textbackslash Hack Hack Hack\textbackslash Whisker\textbackslash Whi-\newline sker\textbackslash obj\textbackslash Release\textbackslash Whisker.pdb} respectively. 

Following the discovery of the \textbf{\textit{tonzking123}} and \textbf{\textit{thobt}} usernames within the \textit{PDB} paths of the compiled malware, an \textit{Open-Source Intelligence} (\textit{OSINT}) investigation was launched to gather further information on the potential actors. Although the association between the usernames found in the malware and the online profiles detailed below is inferred and not confirmed by direct evidence, the unique nature of these identifiers warrants a thorough examination. The following analysis details the findings from various online sources associated with these usernames.

\subsubsection{OSINT Analysis: tonzking123}
\begin{figure}[H]
    \centering
    \frame{\includegraphics[width=0.7\linewidth]{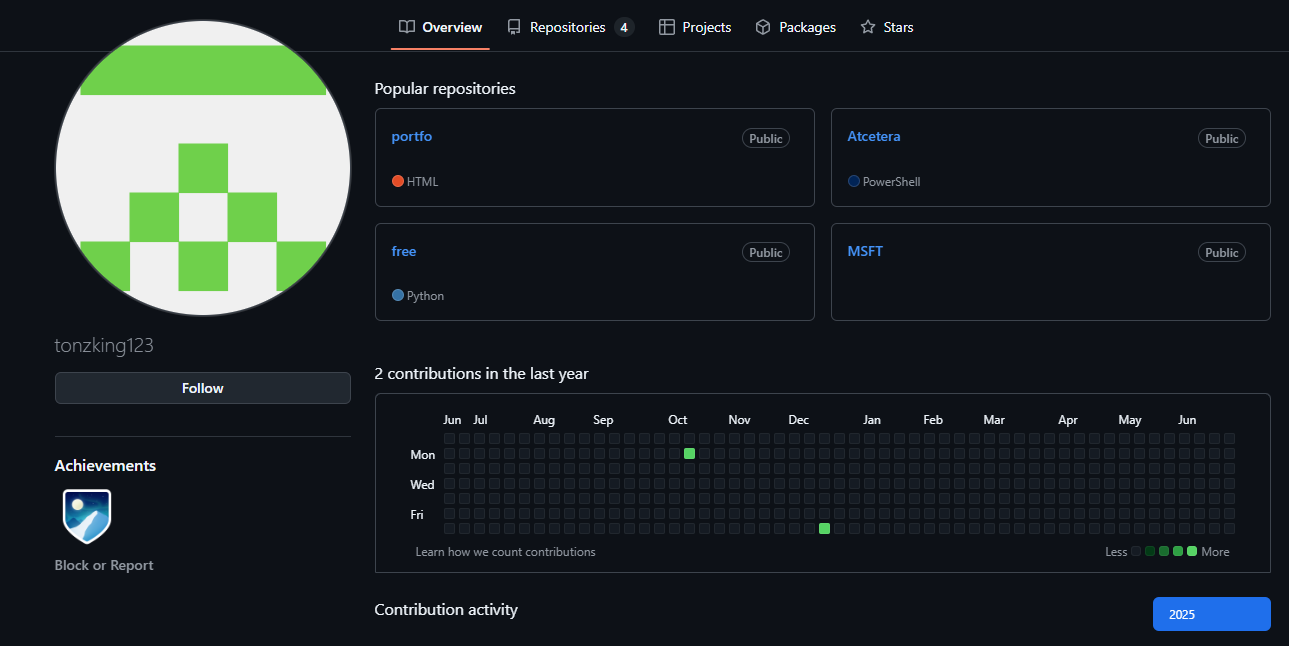}}
    \caption{tonzking123 \textit{Github} account}
    \label{fig:github1}
\end{figure}

The username \textit{tonzking123} appears across several platforms, suggesting a consistent online identity. A \href{https://github.com/tonzking123}{GitHub account} with this username (Fig. \ref{fig:github1}), though not highly active, demonstrates an interest in software development. The profile was created several years ago but shows limited public activity, which could indicate that its primary use is for private repositories or infrequent contributions.

\begin{figure}[H]
    \centering
    \frame{\includegraphics[width=0.8\linewidth]{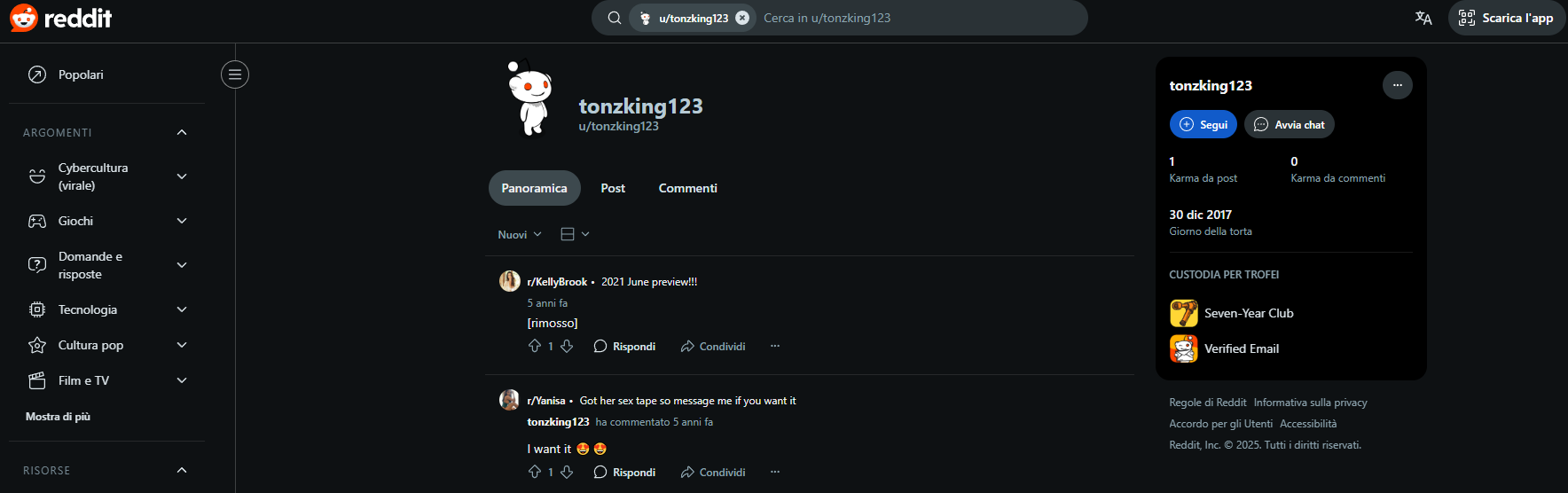}}
    \caption{tonzking123 \textit{Reddit} profile}
    \label{fig:reddit}
\end{figure}

Further investigation uncovered a \href{https://www.reddit.com/user/tonzking123/}{Reddit profile} (Fig. \ref{fig:reddit}). An analysis of the user's posts and comments reveals a range of interests, from discussions on technology and gaming to participation in various niche communities. This activity provides a broader personal context, though it does not directly link to malicious activities.

Analyses carried out on Facebook profiles have shown several different accounts with this name, but no direct association or evidence useful to point out of these specific accounts were found.

\begin{figure}[H]
    \centering
    \frame{\includegraphics[width=0.8\linewidth]{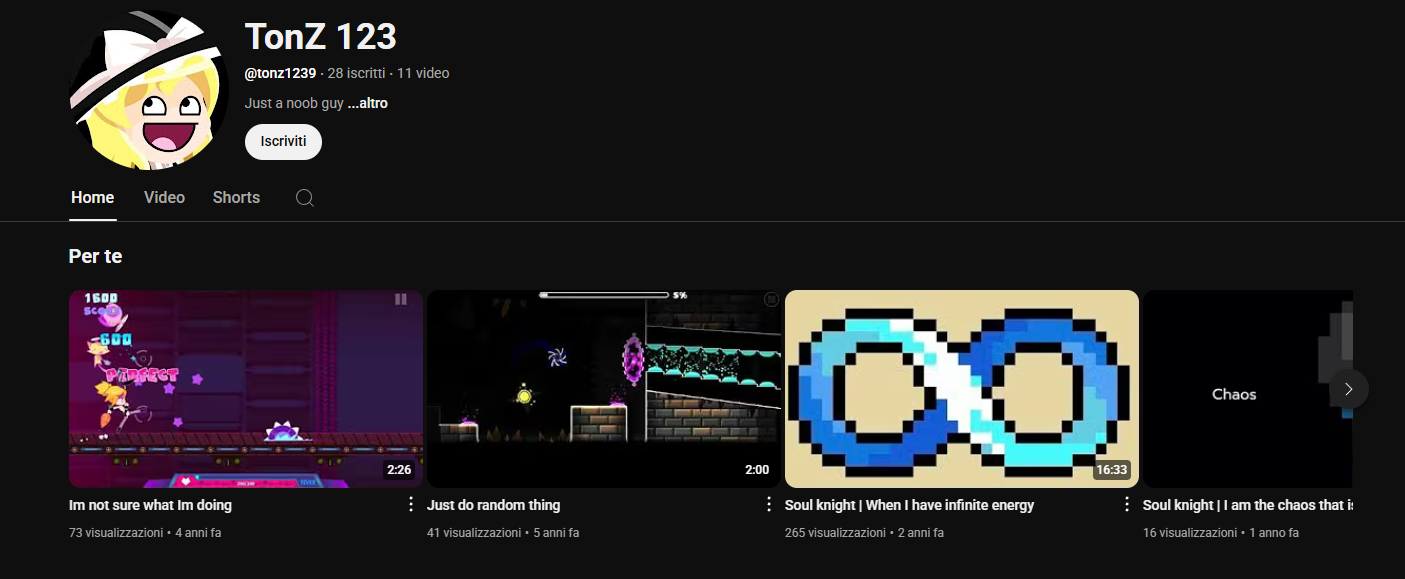}}
    \caption{Tonz123 \textit{Youtube} channel}
    \label{fig:youtube}
\end{figure}

A link to a Google user content URL associated with a \href{https://www.youtube.com/channel/UCPCVwGsRLTwenYclQxTdKtA/posts}{YouTube account} was also found(Fig. \ref{fig:youtube}), further solidifying the presence of this user across Google's services, although the specific content could not be retrieved.

Collectively, these findings paint a picture of an individual with a long-standing and consistent online presence. However, none of the identified accounts can be surely associated with the identified malicious opendir infrastructure.

\subsubsection{OSINT Analysis: thobt}
\begin{figure}[H]
    \centering
    \frame{\includegraphics[width=0.8\linewidth]{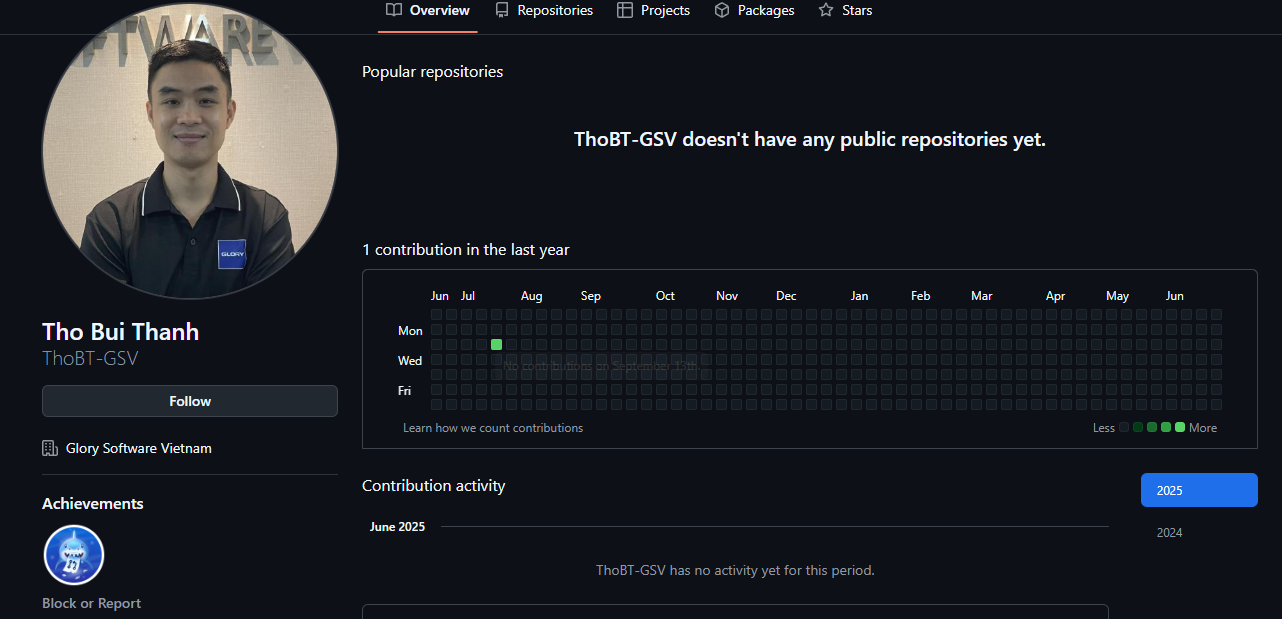}}
    \caption{ThoBT-GSV \textit{Github} profile}
    \label{fig:github2}
\end{figure}

The investigation into the username \textit{thobt} primarily centered on GitHub, where two distinct profiles were identified that suggest a connection to an individual with significant technical expertise. The \href{https://github.com/ThoBT-GSV/}{first GitHub profile} (Fig. \ref{fig:github2}), contains several repositories that align with the tools found in the open directory. This account hosts projects related to cybersecurity and system-level programming, indicating a skillset that would be necessary to compile and modify the malware in question.

\begin{figure}[H]
    \centering
    \frame{\includegraphics[width=0.8\linewidth]{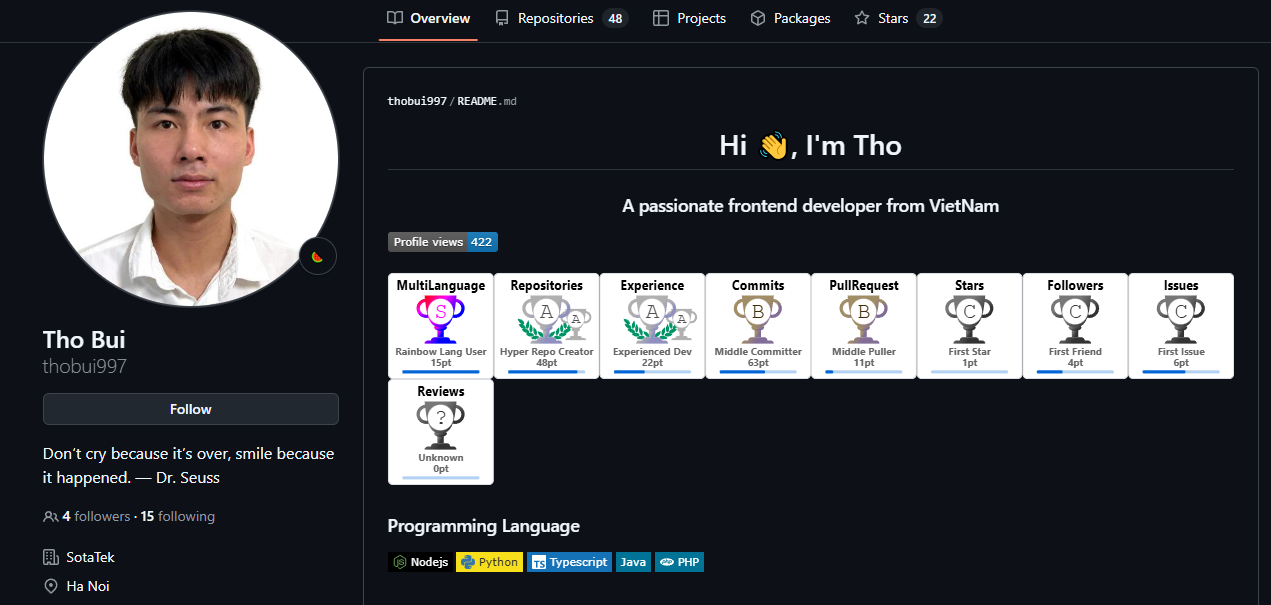}}
    \caption{thobui997 \textit{Github} profile}
    \label{fig:github3}
\end{figure}

The \href{https://github.com/thobui997}{second GitHub account} (Fig. \ref{fig:github3}), appears to be an older or alternative profile for the same individual. The username \textit{thobui} combined with the number $997$ could potentially signify a name and a birth year or another significant number. This profile also features repositories focused on software development and computer science, reinforcing the technical proficiency of the user. The presence of two related \textit{GitHub} accounts could suggest an evolution in the user's online persona or a separation between professional or public projects and more experimental or private work.

While direct attribution cannot be made, the technical evidence from the repositories on these GitHub accounts provides a compelling, albeit inferred, link to the capabilities required to orchestrate the previously analyzed malware deployment.

\subsection{Havoc payloads}
A concise overview of the Havoc malware family—taken out via the thorough treatment of its architecture, lineage and historical campaigns, is referred to the dedicated appendix \ref{havoc}, where those themes are addressed in detail. 

Within the same main exposed folder lay the six Havoc Demons analyzed here: \textit{march.exe}, \textit{https.exe}, \textit{baboon.exe}, \textit{demonx64.exe}, \textit{demon.exe} and \textit{c2\_payload\_aes.exe}. Each file name (except for the encrypted one) followed the attacker’s usual naming convention—short, lower-case words divorced from obvious version strings—suggesting the directory was not intended for public browsing but for automated retrieval by first-stage droppers that receive a hard-coded URL and silently download whichever file is specified in the script.

The co-location of every variant in one place is unlikely to be accidental. The most economical explanation is that the opendir functions as a staging area from which the operator can distribute any build best matches a victim’s profile. The presence of both thirty-two- and sixty-four-bit versions, as well as subtly different resource payloads, implies an attempt at broad compatibility across operating systems and localization settings. By hosting every build together, the attacker simplifies logistics: a single web server can satisfy all requests, and a dropper need supply only the file name to obtain the correct payload. It is equally plausible that the directory doubles as a private version-control snapshot. 

\subsubsection{Demons Static Analysis}
\begin{figure}[H]
    \centering
    \frame{\includegraphics[width=0.7\linewidth]{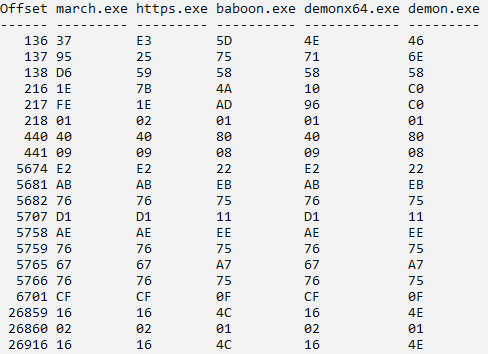}}
    \caption{Differences found in the five main available HAVOC Demons.}
    \label{fig:comparison}
\end{figure}

\begin{figure}[H]
    \centering
    \frame{\includegraphics[width=0.8\linewidth]{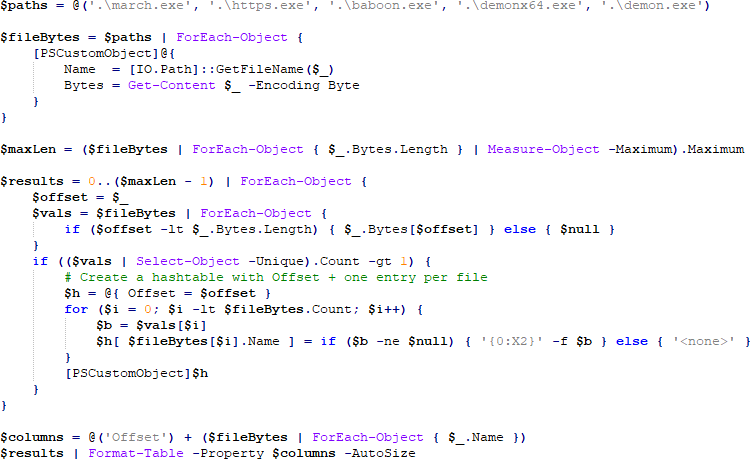}}
    \caption{Code used to detect differences between Demons.}
    \label{fig:codedem}
\end{figure}

The comparative byte-level examination of the five executables reveals a family of binaries that share an overwhelmingly common code base yet diverge in a small, highly patterned set of locations (Fig. \ref{fig:comparison}). Throughout most of the address space the same value recurs across all samples, attesting to a stable core that has remained intact during successive compilations or packer runs (Fig.\ref{fig:codedem}). The few offsets in which inconsistencies appear are therefore particularly instructive, because they expose branch points where a build-time option, a conditional compilation symbol, or a post-packaging patch was applied.

At several lower offsets—for example $136-138$, $216-218$ and $440-441$—pairs or triplets of files alternate predictably between two constant values. A canonical illustration is the flip between \textit{0xAB} and \textit{0xEB} at offset $5681$, which recurs verbatim at $5758$, $5765$, $6701$ and other addresses. Wherever \textit{0xAB} is found in \textit{march.exe}, \textit{https.exe} and \textit{demonx64.exe}, its complement \textit{0xEB} appears in \textit{baboon.exe} and \textit{demon.exe}. Such symmetry is characteristic of a compile-time constant (perhaps a checksum seed, an encryption key byte or a version marker) that can be toggled by a build script. In practice it partitions the corpus into two coherent sub-groups: a first group comprising march.exe, \textit{https.exe} and \textit{demonx64.exe} and a second consisting of \textit{baboon.exe} and \textit{demon.exe}. Within each group intra-binary variation is minimal; between groups, the same handful of offsets carry the opposite nibble pattern every time, confirming the presence of a deliberate variant flag rather than random corruption.

The structural affinity between \textit{march.exe}, \textit{https.exe} and \textit{demonx64.exe} is further reinforced by the observation that \textit{demonx64.exe}—despite its 64-bit architecture—tracks the byte pattern of the other two almost without exception. This suggests that the 64-bit build was produced with the same source and the same configuration switches, then subjected to a routine cross-compilation step. Conversely \textit{baboon.exe} and \textit{demon.exe} form their own pair: when one departs from the majority byte, the other follows it with near perfect fidelity. That mutual dependence implies that both were generated by the same \textit{toolchain} invocation, but under a flag set distinct from that used for the remaining trio.

A second locus of interest emerges above the ninety-eight-thousand-byte mark, where \textit{baboon.exe} begins to diverge dramatically by introducing long sequences of non-zero, often ASCII-printable bytes while most of the other binaries remain padded with zeros or default filler. The insertion of text-like data at this late offset is typical of an embedded resource region—plain-text configuration, hard-coded URLs, or artifacts of a custom packer. Its exclusive presence in \textit{baboon.exe} (and, sporadically, \textit{demon.exe}) strengthens the conclusion that Variant B was packed with additional runtime data or diagnostic strings that the Variant A \textit{toolchain} suppresses.

These patterns can be leveraged operationally (Tab. \ref{tab:havoc_variants}). A detector need examine only two or three signature offsets—for instance $5681$ and $6701$—to classify an unknown specimen unambiguously into Variant A or Variant B. Moreover, the extended ASCII block in Variant B offers a fertile source of static indicators: extracting and decoding those strings is likely to yield mutex names, command-and-control hostnames or file system paths that do not appear in the leaner Variant A builds.

Furthermore, a recurring and notable string, \textit{not havoc}/\textit{NotHavoc} was discovered embedded within the binaries (Tab. \ref{tab:havoc_variants}). While this specific string is not a known default artifact of the public \textit{Havoc C2 framework}, its deliberate inclusion strongly suggests a conscious effort by the developer to create a distinction, likely for the purposes of misdirection. This tactic is conceptually similar to historical precedents in malware naming and analysis, most famously illustrated by the distinction between the \textit{Petya} and \textit{NotPetya} malware. Making a comparison to this case provides a valuable framework for understanding the potential intent behind such a finding.

The \textit{Petya}/\textit{NotPetya} case serves as a critical example of how two seemingly related pieces of malware can have fundamentally different purposes. The original \textit{Petya} was a financially motivated ransomware, designed to encrypt a victim's \textit{Master File Table} (\textit{MFT}) and extort a payment in exchange for a working decryption key. It was, by all accounts, a criminal enterprise tool\footnote{https://www.fortinet.com/it/corporate/about-us/petya-ransomware}. In 2017, a new threat emerged that shared surface-level characteristics with \textit{Petya}, including a similar ransom note. However, security researchers quickly identified crucial technical differences. This new variant, dubbed \textit{NotPetya} by Kaspersky Lab to emphasize this distinction, was not ransomware but a destructive cyberweapon\footnote{https://www.kaspersky.com/blog/new-ransomware-epidemics/17314/}. Its encryption routine was intentionally irreversible, and its primary goal was not financial gain but widespread, catastrophic data destruction, particularly targeting \textit{Ukrainian} entities. The name \textit{NotPetya} was thus an external label applied by the security community to prevent confusion and accurately classify the threat as a wiper, not ransomware.

Applying this precedent to the discovery of the \textit{not havoc}/\textit{NotHavoc} string allows for a correct and insightful assumption. The key difference is the origin of the label. In the case of \textit{NotPetya}, the name was applied by defenders after forensic analysis. Here, the string is an internal artifact, placed by the malware author themselves. This represents a proactive attempt to manipulate the analytical process and control the narrative. By embedding this string, the threat actor is likely engaging in a form of preemptive false-flagging or deception.

The motivations for such an action can be manifold. It may be an attempt to mislead an analyst into believing the tool is merely an amateur or modified version of the well-known \textit{Havoc framework}, thereby misattributing the activity to a different group known to use the public version. Conversely, it could imply that the tool is a significantly more advanced or custom-built framework that is merely masquerading as \textit{Havoc}, with the string serving as an ironic internal nod to its true, distinct nature. This act of embedding a deceptive identifier is a clear attempt to control the narrative of an intrusion, complicating attribution and forcing analysts to question the provenance of the tools they are examining. However, as per the current analysis the demons' behavior seems to not get too far away from how literature describes this threat.

\begin{table}[htbp]
  \centering
  \small
  \setlength\extrarowheight{2pt}
  \begin{tabular}{|
     >{\centering\arraybackslash}m{3cm}|
     >{\centering\arraybackslash}m{4cm}|
     >{\centering\arraybackslash}m{3cm}|
     >{\centering\arraybackslash}m{4cm}|
  }
    \hline
    \textbf{File} & 
    \textbf{Compilation Date} & 
    \textbf{Imports} & 
    \textbf{Useful Strings} \\ 
    \hline

    demon.exe &
    Tue Dec 10 16:37:26 2024 UTC &
    N/A &
    \parbox[c][4cm][c]{\linewidth}{%
      \centering
      \texttt{Mozilla/5.0 (Windows NT 6.1; WOW64) AppleWebKit/537.36} \\ 
      \rule{\linewidth}{0.4pt}\\
      \texttt{not havoc}%
    } \\
    \hline

    baboon.exe &
    Tue Dec 10 17:07:41 2024 UTC &
    N/A &
    \texttt{NotHavoc} \\
    \hline

    demonx64.exe &
    Tue Dec 10 16:50:22 2024 UTC &
    N/A &
    \texttt{Mozilla/5.0 (Windows NT 6.1; WOW64) AppleWebKit/537.36 (KHTML, like Gecko) Chrome/96.0.4664.110 Safari/537.36} \\
    \hline

    https.exe &
    Wed Dec 11 05:40:51 2024 UTC &
    N/A &
    \texttt{Mozilla/5.0 (Windows NT 6.1; WOW64) AppleWebKit/537.36 (KHTML, like Gecko) Chrome/96.0.4664.110 Safari/537.36} \\
    \hline

    march.exe &
    Sun Mar 16 09:09:11 2025 UTC &
    N/A &
    \texttt{Mozilla/5.0 (Windows NT 6.1; WOW64) AppleWebKit/537.36 (KHTML, like Gecko) Chrome/96.0.4664.110 Safari/537.36} \\
    \hline

    c2\_payload\_aes.exe &
    Sun Apr 27 07:20:56 2025 UTC &
    ADVAPI32.dll, KERNEL32.dll, msvcrt.dll &
    \texttt{DHANUSHCODE56, DHANUSHKEY1} \\
    \hline

  \end{tabular}
  \caption{Summary of interesting strings found in the available HAVOC Demons.}
  \label{tab:havoc_variants}
\end{table}

In sum, the diff manifests a controlled, deterministic divergence rather than stochastic noise. All five executables descend from the same parent code base; the differences map to a small set of bytes that are toggled coherently, pointing to two distinct compilation or packing profiles. One profile produces the march.exe lineage, light in resources and internally consistent across 32- and 64-bit builds. The other profile generates the baboon.exe lineage, distinguished by the complementary constant pattern and by the injection of a late resource section. Recognizing and tracking these invariant offsets affords a reliable method of clustering future samples and of uncovering embedded configuration data that may otherwise stay hidden.

\begin{figure}[H]
    \centering
    \frame{\includegraphics[width=1\linewidth]{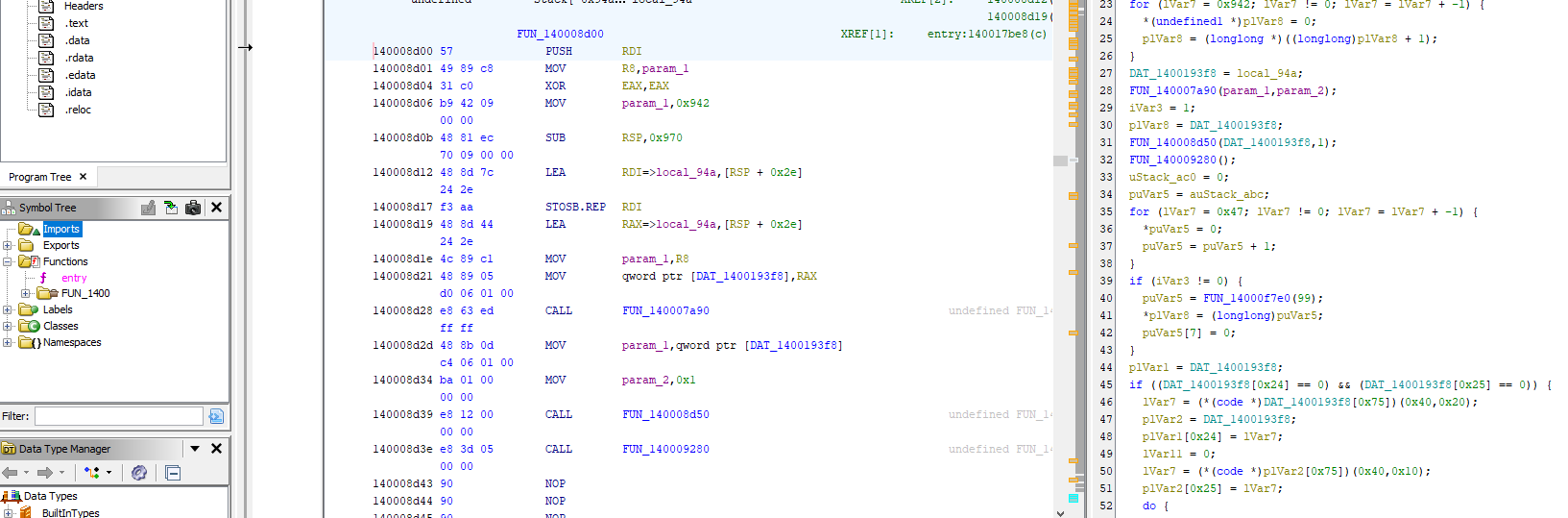}}
    \caption{Demons have no imports or explicit \textit{WinAPI} call.}
    \label{fig:imports}
\end{figure}

A critical design feature of the \textit{Havoc C2} framework's agents is its systematic avoidance of explicit, statically defined imports (as found for \textbf{\textit{march.exe}}, \textbf{\textit{https.exe}},  \textbf{\textit{demonx64.exe}}, \textbf{\textit{demon.exe}} and \textbf{\textit{baboon.exe}}). This is a deliberate and sophisticated defense evasion technique aimed squarely at complicating, if not entirely neutralizing, traditional static analysis methods. Instead of declaring necessary \textit{Windows API} functions in its file header, the Havoc demon resolves their addresses dynamically at runtime, a method that fundamentally alters how the malware appears to security analysts and their automated tools (Fig. \ref{fig:imports}). This behavior can be observed in all of the five non encrypted executable mentioned before.

In a standard executable, the functions it intends to use from system libraries like \textit{kernel32.dll} or \textit{ntdll.dll} are listed within its \textit{Import Address Table} (\textit{IAT}). Security analysts rely heavily on inspecting the \textit{IAT} as a first step to quickly understand a program's capabilities. For instance, the presence of functions like \textit{VirtualAlloc}, \textit{CreateRemoteThread}, and \textit{WriteProcessMemory} in the \textit{IAT} would immediately flag a file as suspicious and capable of process injection. Havoc circumvents this entirely. When statically analyzed, a Havoc demon's \textit{IAT} is conspicuously sparse, often showing only the most basic functions required to load the program itself, giving no indication of its true malicious intent.

To achieve this, the demon employs dynamic \textit{API} resolution. Upon execution, instead of relying on the \textit{PE loader} to link its functions, the agent actively finds the memory addresses of the functions it needs. This is typically accomplished by first obtaining the base address of a core system library, such as \textit{ntdll.dll}, by traversing the \textit{Process Environment Block} (\textit{PEB}). Once the library is located in memory, the demon can then parse its \textit{Export Address Table} to find specific functions. However, to further obscure its operations, \textit{Havoc} does not search for function names directly as plaintext strings. Instead, as confirmed by multiple security researchers, it uses a custom hashing algorithm (a variant of \textit{DJB2}) to calculate the hash of the required function name (e.g. \textit{NtAllocateVirtualMemory}) and then compares this against the hashes of all function names exported by the library until it finds a match. This hashing technique prevents analysts from simply searching the binary for suspicious \textit{API} name strings.

Furthermore, \textit{Havoc} elevates this technique by implementing indirect \textit{syscalls}. Rather than calling the resolved \textit{API} function directly, the demon retrieves the function's unique \textit{syscall} number. It then executes the call instruction directly, effectively bypassing user-mode \textit{API} hooking employed by many \textit{Endpoint Detection and Response} (\textit{EDR}) solutions. By combining dynamic \textit{API} resolution through hashing with the execution of indirect \textit{syscalls}, the \textit{Havoc} demon successfully severs the explicit links between its code and its malicious actions, forcing analysts to perform far more complex and time-consuming dynamic analysis and in-depth reverse engineering to uncover its true functionality.

\begin{figure}[H]
    \centering
    \frame{\includegraphics[width=1\linewidth]{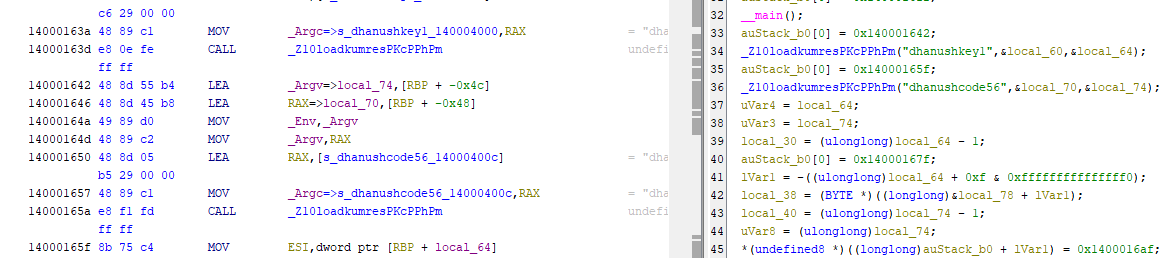}}
    \caption{Snippet of the first part of the main routine of \textit{c2\_payload\_aes.exe}}
    \label{fig:crypto}
\end{figure}

On the other hand, the \textit{main} routine of the \textbf{\textit{c2\_payload\_aes.exe}} program functions as a sophisticated in-memory loader and execution stage for an encrypted payload, a technique commonly referred to as a \textit{stager} or \textit{dropper} (Fig. \ref{fig:crypto}). The code meticulously avoids writing the final-stage payload directly to disk, instead orchestrating its decryption and execution entirely within the process's memory space to evade static detection by security products.

The process commences by invoking a custom function, \textit{\_Z10loadkumresPKcPPhPm}, twice in succession. This function appears to load two distinct resources embedded within the binary, identified by the string tags \textit{dhanushkey1} and \textit{dhanushcode56}. Judging by these identifiers and the subsequent operations, \textit{dhanushkey1} is almost certainly the decryption key, likely for the AES algorithm mentioned in its name, while \textit{dhanushcode56} represents the main encrypted payload or \textit{demon}. Immediately after loading, the code copies both the key and the encrypted payload from their initial locations into newly allocated buffers on the stack. This step serves as a form of obfuscation, moving the critical components to transient memory locations.

\begin{figure}[H]
    \centering
    \frame{\includegraphics[width=1\linewidth]{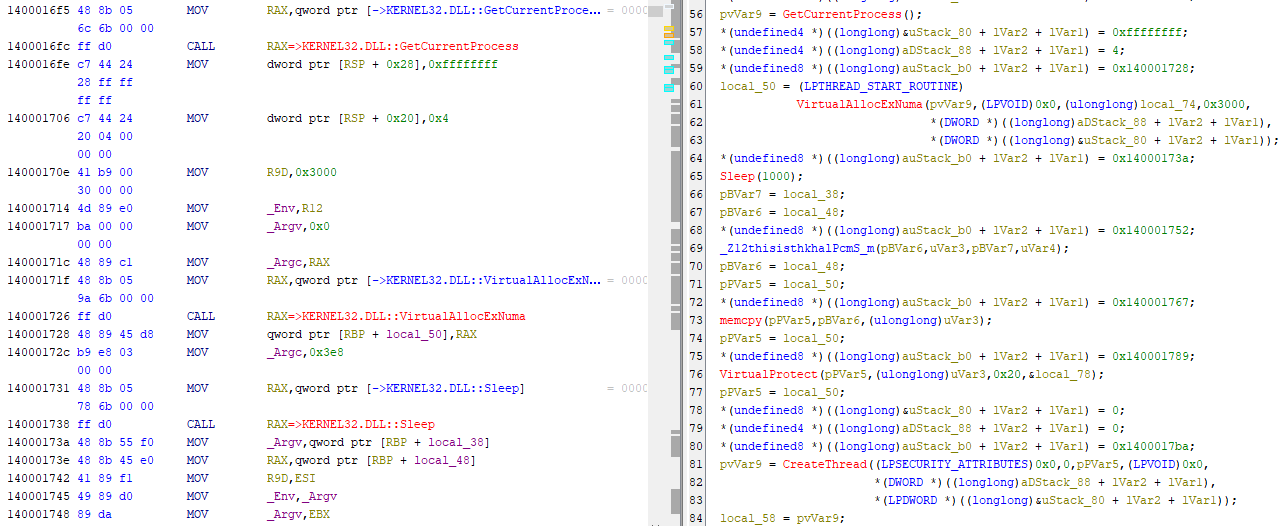}}
    \caption{Shellcode injection instructions in \textit{c2\_payload\_aes.exe}}
    \label{fig:crypto2}
\end{figure}

Following this preparation, the program initiates a classic process injection sequence (Fig. \ref{fig:crypto2}). It first allocates a new region of memory within its own process space using \textit{VirtualAllocExNuma}, carefully specifying the size to match that of the encrypted payload and setting the initial memory protection to \textit{PAGE\_READWRITE}. After a brief one-second pause, likely an elementary anti-sandboxing measure, the code calls another internal, obfuscated function named \textit{\_Z12thisisthkhalPcmS\_m}. This routine is passed the pointers to both the encrypted payload and the key, and it is here that the in-memory decryption of the payload occurs. Once decrypted, the plaintext shellcode is copied from its stack buffer into the newly allocated memory region. To make this region executable, a call to VirtualProtect is made, altering its permissions from \textit{PAGE\_READWRITE} to \textit{PAGE\_EXECUTE\_READ}. With the payload now decrypted and residing in an executable memory segment, the program launches it by creating a new thread of execution via \textit{CreateThread}, with the starting address pointing directly to the payload. Finally, the main thread enters a wait state using \textit{WaitForSingleObject}, ensuring the loader process remains active and persists until the malicious thread has completed its execution.

In sum, the open directory provided a convenient, one-stop repository for every active variant of the \textit{HAVOC Demon}. Its contents reveal an attacker who values agility—using multiple compilations to test detection thresholds or target different environments—but who is willing to trade secrecy for operational simplicity, betting that the directory will remain unnoticed for as long as the campaign remains profitable.

\subsubsection{Demons Dynamic Analysis}
\begin{figure}[H]
    \centering
    \frame{\includegraphics[width=0.8\linewidth]{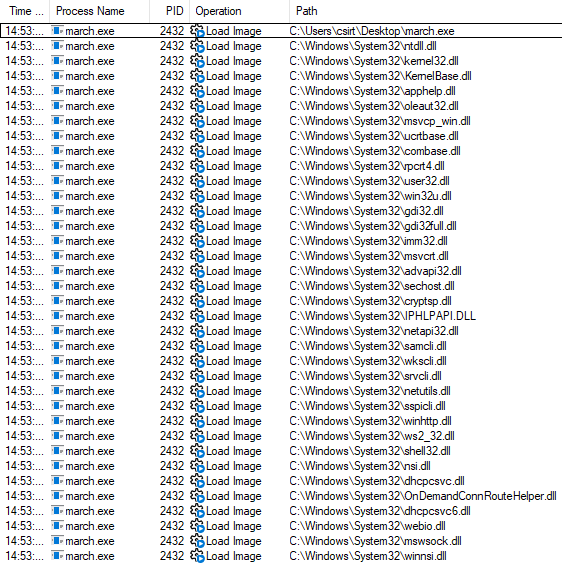}}
    \caption{DLLs loaded by march.exe at runtime.}
    \label{fig:marchdll}
\end{figure}

A detailed examination of the memory dump of the \textbf{\textit{march.exe}} process provides a comprehensive profile of a sophisticated implant. The technical evidence points toward a full-featured backdoor designed for stealth, persistence, and secure command-and-control communications. A critical indicator of compromise was immediately apparent in the form of a hardcoded network endpoint, \textit{52.230.23[.]114:8443}, which serves as the malware's \textit{C2} server. The mechanism for this communication is strongly suggested by the presence of strings referencing \textit{winhttp.dll} and specific \textit{WinHTTP API} functions such as \textit{WinHttpConnect} and \textit{WinHttpOpenRequest} (Fig. \ref{fig:marchdll}). This indicates that the implant almost certainly utilizes the HTTP/S protocol to establish its C2 channel, a common choice for blending in with legitimate network traffic.

\begin{figure}[H]
    \centering
    \frame{\includegraphics[width=1\linewidth]{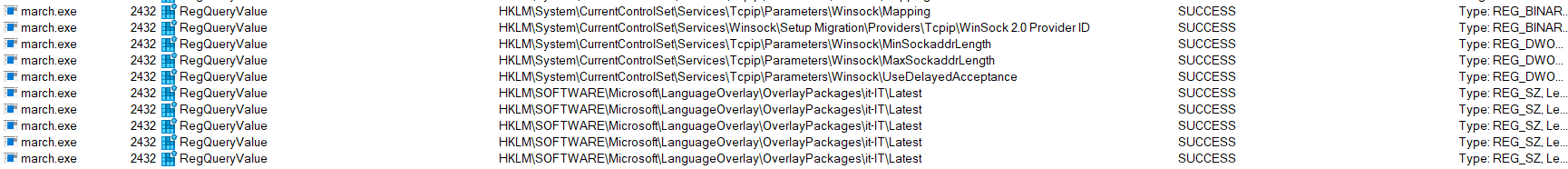}}
    \caption{Example of registry keys queried by \textit{march.exe}.}
    \label{fig:regs}
\end{figure}

On the infected host, the malware demonstrates extensive interaction with the Windows operating system at a low level, revealing its capabilities for persistence, defense evasion, and system reconnaissance. Its behavior show numerous queries and manipulations of the Windows Registry (Fig. \ref{fig:regs}), most notably targeting the \textit{\textbackslash REGISTRY\textbackslash MACHINE\textbackslash \newline SOFTWARE\textbackslash Microsoft\textbackslash Windows NT\textbackslash CurrentVersion\textbackslash Image File Execution Options} key. This is a well-documented technique for achieving persistence by hijacking the execution flow of legitimate system processes. Furthermore, the malware exhibits an awareness of the security posture of the host by querying \textit{\textbackslash REGISTRY\textbackslash MACHINE\textbackslash SOFTWARE\textbackslash \newline Policies\textbackslash Microsoft\textbackslash Windows\textbackslash Safer\textbackslash CodeIdentifiers}, suggesting an attempt to fingerprint to possible later bypass software restriction policies like \textit{AppLocker}.

The implant's functionality is further illuminated by its use of cryptographic libraries. The loading of \textit{bcrypt.dll} and \textit{cryptsp.dll}, coupled with function calls related to the \textit{CryptoAPI} such as \textit{CPAcquireContext}, \textit{CPEncrypt}, and \textit{CPGenKey}, confirms that the malware employs strong encryption. This cryptographic capability is likely used for two primary purposes: to protect its \textit{C2} communications from network inspection and to potentially encrypt data or its own components on the host. The combination of these features—secure \textit{C2} communication, advanced persistence mechanisms, and defense evasion techniques—characterizes march.exe as a potent threat, warranting immediate incident response focused on the identified network and host-based indicators.

\begin{figure}[H]
    \centering
    \frame{\includegraphics[width=1\linewidth]{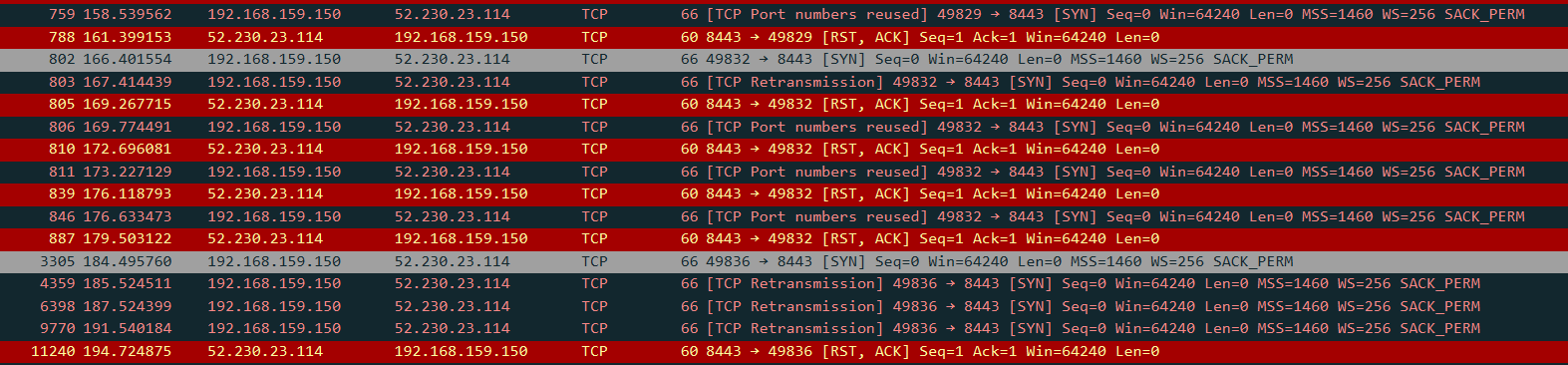}}
    \caption{\textit{C2} Beaconing activity directed to \textit{52.230.23[.]114:8443}.}
    \label{fig:c2bec}
\end{figure}

Upon a comparative analysis of the artifacts from the \textbf{\textit{march.exe}} and \textbf{\textit{baboon.exe}} executables, it is evident that both samples are variants derived from the same malware family, exhibiting a shared core architecture and operational infrastructure. The most definitive link is the use of the identical command-and-control server, hardcoded as \textit{52.230.23[.]114}, indicating that both implants are designed to communicate with the same backend (Fig. \ref{fig:c2bec}). This foundational similarity is further reinforced by a common set of imported libraries, including \textit{winhttp.dll} for HTTP-based \textit{C2} communications and cryptographic modules like \textit{bcrypt.dll} and \textit{cryptsp.dll} for secure data exchange.

Despite their common origin, notable differences emerge that point to an evolution or specialization of the implant. \textit{baboon.exe} presents a broader range of strings related to the Windows graphical user interface, including references to the \textit{Control Panel} and various GUI class names like \textit{ComboBox} and \textit{ListBox}. This implies a more complex functionality that may involve interacting with or manipulating GUI elements, a feature not apparent in \textit{march.exe}. In synthesis, while both executables share a definitive lineage and \textit{C2} infrastructure, \textit{baboon.exe} represents a more advanced and customized iteration, demonstrating greater environmental awareness and potentially more sophisticated methods of user or system interaction. Moreover, the remaining non-encrypted demons behave similarly to \textit{march.exe}.

\begin{figure}[H]
    \centering
    \frame{\includegraphics[width=1\linewidth]{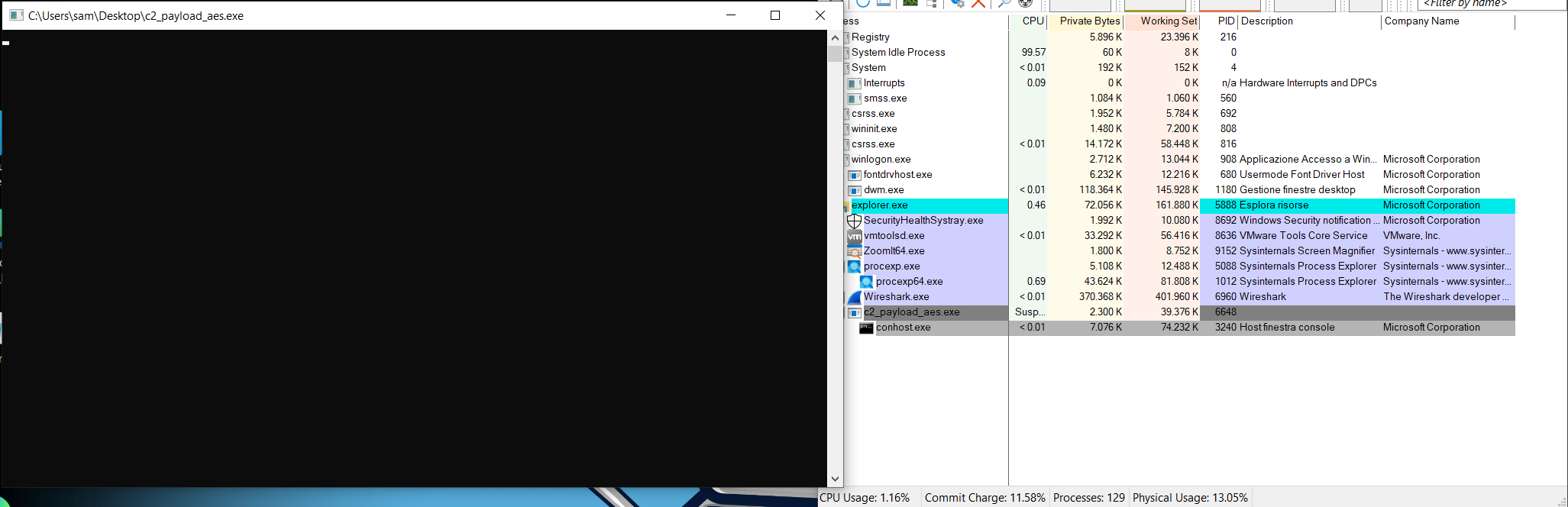}}
    \caption{\textit{conhost.exe} runs with an open windows whenever \textit{c2\_payload\_aes.exe} is executed.}
    \label{fig:conhost}
\end{figure}

\textbf{\textit{c2\_payload\_aes.exe}}, exhibits a striking resemblance to the previously analyzed \textit{march.exe} and \textit{baboon.exe} samples, confirming its origin from the same malware family. The most telling similarity is the continued use of the hardcoded \textit{Command and Control} server at IP address \textit{52.230.23[.]114} on port 8443 , a characteristic shared across all three samples. This strongly suggests a common infrastructure and likely the same threat actor behind these tools. Furthermore, possibly due to some coding overlooking, once the payload runs a \textit{conhost.exe} prompt is shown continuously, alerting the user that the executable is running (Fig. \ref{fig:conhost}).

Functionally, it employs a similar array of Windows libraries for its operations. The continued use of \textit{winhttp.dll} for network communication and \textit{bcrypt.dll} and \textit{cryptsp.dll} for cryptographic functions indicates a consistent methodology for command and control as well as data protection. \textit{bcrypt.dll} and \textit{cryptsp.dll }libraries are essential for implementing AES encryption in a Windows environment. While all three samples exhibit similar host-based indicators and rely on a core set of Windows libraries, the \textit{c2\_payload\_aes.exe} sample provides a more specific insight into the cryptographic methods employed by this malware family. This suggests a potential evolution or specialization in the malware's design, with this particular variant emphasizing the use of AES for secure communication and data protection. The presence of detailed user and environmental information in "baboon.exe" but not in \textit{c2\_payload\_aes.exe} might also indicate different stages of attack or different target profiles for each variant.

\newpage
\section{Defensive Countermeasures}
On endpoints, script-block logging, AMSI telemetry and advanced ETW providers shoudl all be enabled; the public Sigma and YARA sets supplied with the report already detect the loader’s reflective-DLL stub and the obfuscated phishing page. Hardening \textit{\%LOCALAPPDATA\%\textbackslash Temp} so that users may create but not delete files breaks the \textit{PowerShell} stage outright and turns an otherwise fileless infection into a benign crash. \textit{AppLocker} or \textit{WDAC} policies that forbid \textit{rundll32} + \textit{netsh} + \textit{esentutl} chains under non-administrative tokens disrupt the attacker’s post-exploitation toolkit documented in earlier campaigns. Within \textit{Active Directory}, continuous auditing of\textit{ msDS-KeyCredentialLink} changes (Event \textit{4662}, \textit{GUID 6f9675b5-…}) is mandatory: the bundled \textit{Whisker} binary silently injects shadow credentials that survive password resets and grant full domain control. Finally, because recent \textit{Havoc} builds tunnel over \textit{Microsoft Graph} or \textit{SharePoint}, defenders must extend \textit{TLS} inspection to sanctioned \textit{SaaS} domains and apply behavioral baselines—for example flagging OneDrive uploads that carry high-entropy blobs at regular beacon intervals. When these layers are combined—patched perimeter, protocol anomaly detection, strict script oversight and directory-integrity monitoring—the infrastructure described in this report becomes noisy, fragile and short-lived.

\newpage
% Conclusion
\section{Conclusion}
This investigation has successfully uncovered and detailed the operations of a sophisticated threat actor utilizing a customized malware framework based on the \textit{Havoc C2}. The analysis of multiple malware samples, combined with \textit{OSINT} research, has allowed for the development of a comprehensive picture of the actor's \textit{TTPs}. The consistent use of the same \textit{C2} infrastructure across all samples, along with the unique \textit{not havoc}/\textit{NotHavoc} string, provides a strong basis for attributing future activity to this same group.  The actor's methods, which include fileless malware, dynamic \textit{API} resolution, and the abuse of trusted infrastructure, highlight the challenges faced by modern defenders. However, the analysis also revealed key weaknesses in the actor's tradecraft, such as the fragile nature of their file-dropping mechanism and the static nature of their \textit{C2} infrastructure, which can be exploited for detection and mitigation. The defensive countermeasures outlined in this report provide a clear roadmap for organizations to protect themselves against this specific threat and other, similar adversaries. Ultimately, this case underscores the importance of a defense-in-depth strategy that combines network security, host-based hardening, and proactive threat intelligence to effectively combat the evolving landscape of cyber threats.
\newpage
% Appendix
\appendix
\section{Appendix}
\subsection{Malware Families and Exploited Tools Background}
This section will provide a comprehensive and detailed exploration of each distinct \textit{Malware Family} identified during the analysis phase. The primary objective is to delineate the fundamental characteristics and historical context pertinent to each family, thereby establishing a robust foundational understanding for subsequent discussions.

Each identified \textit{Family} will be examined individually, with a focus on several key aspects. A thorough background will be presented, encompassing their origins, typical propagation methods, and evolutionary trajectory over time. This historical perspective is crucial for understanding the sophistication and adaptation mechanisms employed by these malicious entities. The core technical characteristics defining each family will be meticulously detailed. This includes, but is not limited to, their operational methodologies, payload delivery mechanisms, persistence techniques, command and control (C2) communication protocols, and unique behavioral traits. Furthermore, any notable variants or sub-families that exhibit significant deviations from the main family's profile will be highlighted and discussed to provide a nuanced understanding of their operational diversity.

By systematically dissecting each malware family in this manner, this section aims to illuminate the intricate landscape of contemporary cyber threats, providing readers with the necessary context to appreciate the complexities involved in malware detection, analysis, and mitigation strategies.

\subsubsection{Havoc}\label{havoc}
\begin{figure}[H]
    \centering
    \frame{\includegraphics[width=0.8\linewidth]{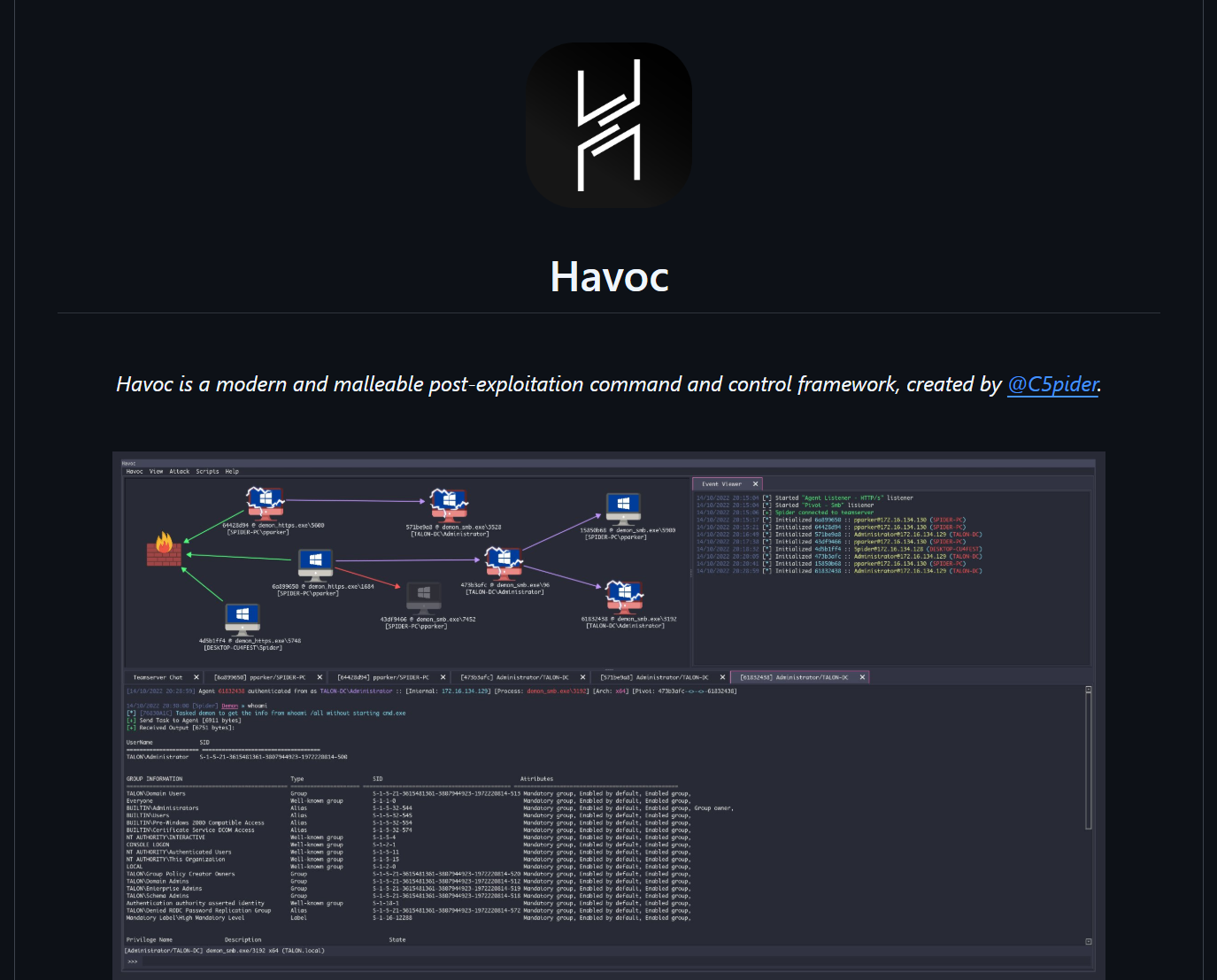}}
    \caption{Havoc Framework's public \href{https://github.com/HavocFramework/Havoc}{GitHub Repository}.}
    \label{fig:1}
\end{figure}

\textbf{\textit{Havoc}} is an open-source \textit{Post-Exploitation} \textit{Command-and-Control framework} released in October 2022 by the developer \textit{C5pider} (Fig. \ref{fig:1}). Its core components are written in \textit{Golang}, \textit{C++}, and \textit{Qt} and the codebase is freely available on \textit{GitHub}, a circumstance that has accelerated both legitimate red-team adoption and malicious repurposing. The framework’s design philosophy mirrors that of \textit{Cobalt Strike} and \textit{Brute Ratel}, but it distinguishes itself through a deliberately minimal public defensive footprint and a strong emphasis on modularity, which together have fostered rapid uptake among threat actors seeking an alternative to heavily signatured tooling\footnote{https://www.immersivelabs.com/resources/blog/havoc-c2-framework-a-defensive-operators-guide}. 

Operationally, \textit{Havoc} implements a classic two-tier architecture comprising a \textit{teamserver}, which orchestrates all listener sockets and tasking queues, and a cross-platform graphical client through which operators issue commands. Compromised hosts run the \textit{Demon} implant, which can be generated as a Windows PE executable, \textit{reflective DLL}, \textit{raw shellcode}, or \textit{shellcode} encapsulated in novel loaders. Callback channels are configurable for \textit{HTTP}, \textit{HTTPS}, and \textit{SMB}, and the implant negotiates encrypted traffic with the \textit{teamserver} at user-defined intervals. The open design allows operators to embed the \textit{C2} channel inside benign protocols or redirect it through reverse proxies without altering core binaries, complicating attribution and sinkholing. 

The Demon agent relies on \textit{AES-256} in \textit{CTR} mode for payload confidentiality, with \textit{keys} and \textit{IVs} negotiated at first contact and stored in the \textit{teamserver}’s \textit{SQLite} database. Each packet is prefixed with the constant \textit{0xDEADBEEF} followed by length and \textit{AgentID} fields, a readily identifiable yet often overlooked \textit{artifact} in raw packet captures. Beacon traffic blends into normal \textit{POST/200} transaction pairs, and inter-beacon dormancy is protected by memory-resident \textit{sleep} techniques—\textit{Foliage}, \textit{Ekko}, or a plain \textit{WaitForSingleObjectEx} loop—that encrypt the entire implant in situ and leverage indirect \textit{syscalls} to evade heuristic scanners on Windows 11\footnote{https://www.zscaler.com/blogs/security-research/havoc-across-cyberspace}. 

The framework’s loader ecosystem reflects current adversary trends toward staged, memory-only delivery. Field observations describe screen-saver downloaders and \textit{BAT-to-EXE} stubs that disable \textit{Event Tracing for Windows}, patch \textit{EtwEventWrite}, and unpack \textit{shellcode} signed with leaked Microsoft certificates. The \textit{shellcode} commonly embeds \textit{KaynLdr}, which reflectively maps a header-stripped \textit{Demon DLL}, resolves \textit{API}s via a modified \textit{DJB2} hashing routine, and immediately initiates the \textit{DemonInit} → \textit{DemonMetaData} → \textit{DemonRoutine} sequence without touching disk. Such loaders also spoof return addresses and tamper with the stack to frustrate dynamic analysis\footnote{https://www.criticalstart.com/new-framework-raising-havoc/}. 

Campaign telemetry collected through 2024-2025 demonstrates \textit{Havoc}’s flexibility in cloud-mediated channels. \textit{FortiGuard} Labs documented a multi-stage phishing operation in which an HTML \textit{ClickFix} lure instructed victims to paste a base-64 PowerShell one-liner\footnote{https://www.fortinet.com/blog/threat-research/havoc-sharepoint-with-microsoft-graph-api-turns-into-fud-c2}; the script downloaded a Python loader from \textit{Microsoft SharePoint}, which in turn deployed a modified Demon compiled to abuse the \textit{Microsoft Graph API}. Instead of direct \textit{POST} beacons, the Demon wrote base-64 job requests into a victim-specific file on SharePoint, then polled the paired one for operator responses, thereby tunneling \textit{C2} traffic through a trusted \textit{SaaS} domain while retaining Havoc’s full command set. 

\textit{Adoption metrics} underscore the framework’s growing threat surface. \textit{ANY.RUN} threat-intelligence telemetry lists \textit{Havoc} among the most active \textit{C2} families as of June 2025, with clusters of servers in North America and Western Europe and a trajectory that correlates with incremental defensive coverage against older frameworks\footnote{https://any.run/malware-trends/havoc/}. This diffusion is aided by the ease with which attackers can compile custom payloads and integrate third-party \textit{Beacon Object Files} (\textit{BOFs}) to extend functionality, from Kerberos attacks to in-memory credential harvesters. 

Practical detection therefore hinges on a convergence of host and network analytics rather than reliance on static indicators. Memory inspection should look for \textit{Threads} in \textit{Wait channel transitions} accompanied by periodic \textit{AES} key material in freshly allocated regions. Network monitoring can fingerprint the \textit{0xDEADBEEF} header and constant packet silhouettes, while threat hunters may extract \textit{Demon encryption keys} from early beacons or the \textit{teamserver} database to decrypt subsequent traffic for full command reconstruction. Because adversaries increasingly encapsulate \textit{Havoc} inside \textit{Graph API} or other cloud endpoints, organizations must enrich \textit{TLS} inspection with contextual domain intelligence and apply anomaly-based detection to outbound \textit{SharePoint} and \textit{OneDrive} traffic. 

In sum, \textit{Havoc} represents the latest evolutionary step in publicly available \textit{C2} tooling: it marries open-source agility with modern evasion primitives, supports a heterogeneous loader ecosystem, and readily blends into \textit{cloud workflows}. The framework’s growing incidence in phishing, cryptocurrency-related intrusions, and state-aligned espionage campaigns attests to its operational maturity.

\subsubsection{Chisel}\label{chisel}
\begin{figure}[H]
    \centering
    \frame{\includegraphics[width=0.65\linewidth]{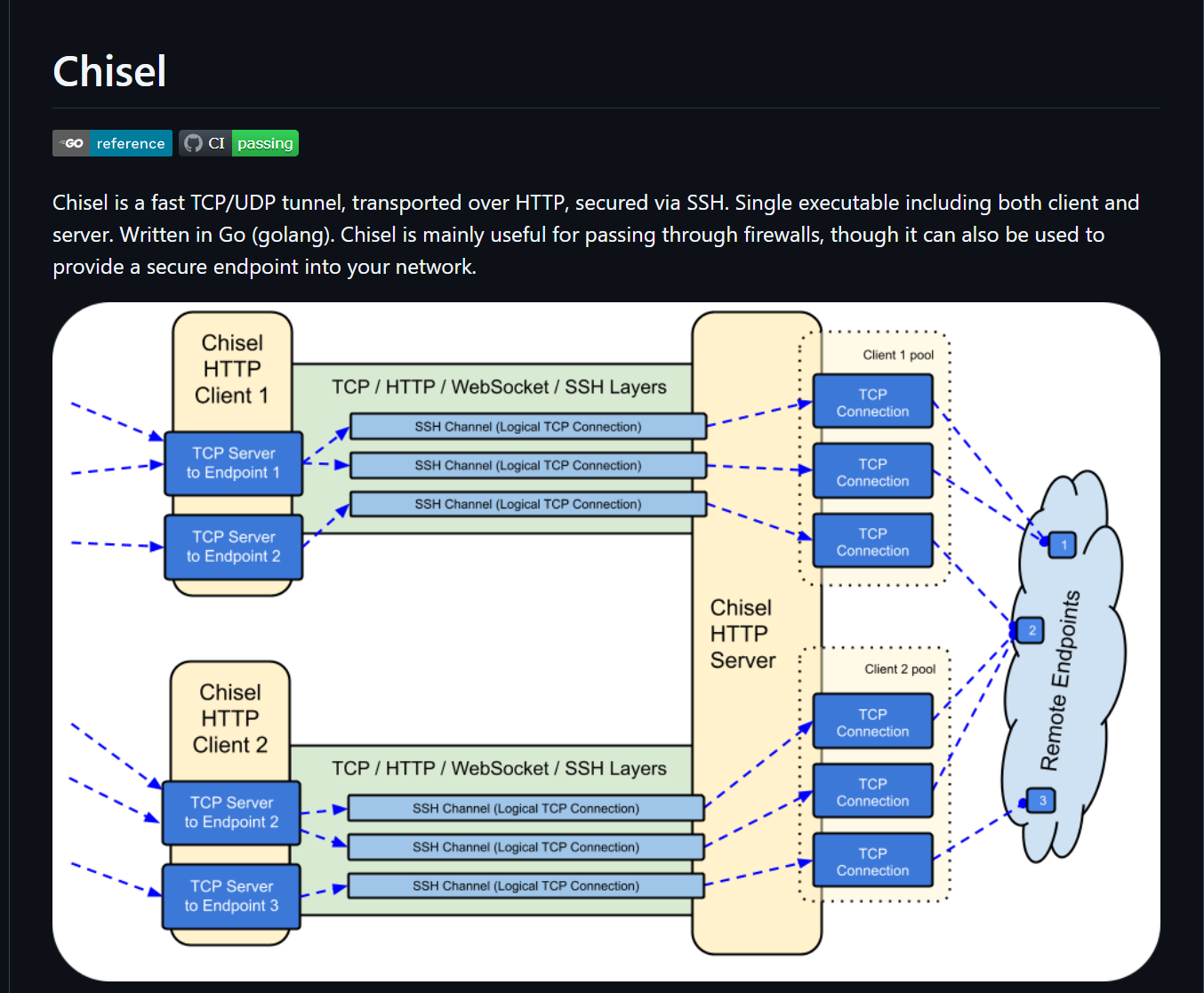}}
    \caption{Chisel public \href{https://github.com/jpillora/chisel}{GitHub Repository}.}
    \label{fig:2}
\end{figure}

\textbf{\textit{Chisel}} is an open-source tunneling utility authored by the developer \textit{jpillora} and written almost entirely in \textit{Go}, which makes it trivially cross-compiled for \textit{Windows}, \textit{Linux}, \textit{macOS} and the major \textit{BSD} derivatives\footnote{https://github.com/jpillora/chisel} (Fig. \ref{fig:2}). 

In its default configuration a single static binary simultaneously houses both client and server logic, enabling users to expose arbitrary \textit{TCP} or \textit{UDP} services through a covert channel that is carried inside ordinary \textit{HTTP} transactions and upgraded to a \textit{full-duplex WebSocket} stream. Because the transport is wrapped in the Go standard-library implementation of SSH, every byte that traverses the tunnel is protected by the cipher suites available in crypto/ssh—typically \textit{AES-CTR} or \textit{ChaCha20-Poly1305}—while the server’s \textit{ECDSA} host key is exchanged in the handshake to provide mutual authentication and resistance to man-in-the-middle interference. The project’s \textit{README} foregrounds these strengths, notably its single-file deployment model, SSH-backed encryption, and the ability to multiplex multiple tunnels across a single TCP connection without material performance loss.

At run time the server listens on an arbitrary \textit{TCP} port, accepts an \textit{HTTP} \textit{GET} bearing the Upgrade: \textit{websocket} header, and then switches protocols with a \textit{101} response; from this point onward the payload is opaque binary data indistinguishable—in length and timing—from benign \textit{WebSocket} traffic. Subsequent reconnections use an exponential-back-off algorithm, and clients can request either forward or reverse port-forwarding, or spawn an embedded \textit{SOCKS5} proxy so that any application capable of speaking \textit{SOCKS} can pivot through the tunnel. The approach is intentionally minimalist: it dispenses with separate key-exchange daemons or layered \textit{stagers}, preferring a direct \textit{GO}→\textit{GO} pipe that reduces operational overhead and frustrates \textit{deep-packet-inspection} systems that do not terminate \textit{TLS}. Independent practitioners have confirmed that once the \textit{WebSocket} session is established, bytes flow bidirectionally without further \textit{HTTP} framing, an efficiency gain that explains Chisel’s popularity among red-team operators as a drop-in replacement for older, header-heavy tools such as \textit{Crowbar}\footnote{https://hackers-arise.com/pivoting-within-the-network-getting-started-with-chisel/}.

The same attributes that appeal to penetration testers have propelled Chisel into mainstream adversary toolchains. The joint \textit{CISA-FBI} advisory on \textit{BlackSuit} ransomware notes that affiliates routinely deploy Chisel alongside \textit{Cloudflared} and \textit{OpenSSH} to sculpt covert paths out of segmented networks, thereby shielding command servers and exfiltration endpoints from perimeter logging\footnote{https://www.cisa.gov/news-events/cybersecurity-advisories/aa23-061a}. \textit{Cyble}’s analysis of a 2024 multi-stage \textit{PowerShell} campaign demonstrates a canonical workflow in which an encoded one-liner retrieves the \textit{Chisel client}, connects to an attacker-controlled server over HTTPS port 443, establishes a SOCKS5 listener on the infected host, and then funnels Nmap and RDP traffic through the newly created conduit to reach systems that would otherwise be unreachable from the public Internet. Ransomware operators are not alone: \textit{BleepingComputer}’s February 2025 coverage of the \textit{BadPilot} espionage program shows \textit{Sandworm}’s \textit{Seashell Blizzard} subgroup pairing \textit{Chisel} with \textit{Rclone} to siphon gigabytes of industrial telemetry while eluding egress filters that block uncommon protocols but leave \textit{WebSocket} over \textit{TLS} untouched\footnote{https://www.bleepingcomputer.com/news/security/badpilot-network-hacking-campaign-fuels-russian-sandworm-attacks/}.

The forensic footprint of a live \textit{Chisel} session is subtle yet discernible. On disk analysts may encounter a transient binary named \textit{chisel.exe}, client, or a randomly generated string, often executed with flags such as client \textit{--reverse} \textit{--socks5} or embedded in a \textit{PowerShell Invoke-Expression} chain. In memory, the \textit{Go} runtime’s characteristic heap layout and symbol table can betray the process to a trained eye, though many threat actors strip symbols to impede static inspection. Network traces captured before \textit{TLS} negotiation reveal an \textit{HTTP GET} whose \textit{User-Agent} defaults to \textit{chisel/X.Y.Z} unless explicitly overridden; thereafter, the traffic assumes the cipher-suite-determined entropy profile of \textit{SSH} and is multiplexed inside the \textit{WebSocket} frame structure, typically keeping the \textit{TCP} window full to maximize \textit{throughput—behavior} that anomalously inflates byte counts for what nominally appears to be an interactive web connection.

Finally, it is worth disambiguating the open-source tunneler from \textit{Infamous Chisel}, a \textit{Sandworm Android backdoor} documented by \textit{CISA} in \textit{2023}; despite partial name overlap, the latter is a bespoke malware platform unrelated to \textit{jpillora}’s project. The coincidence, however, illustrates \textit{Chisel}’s standing in the attacker lexicon: the term itself has become shorthand for covert, firewall-agnostic transport, whether realized through the \textit{Go} binary examined here or through bespoke code that imitates its salient techniques. In an environment where encrypted \textit{WebSocket} traffic over port 443 is operationally indistinguishable from everyday SaaS usage, \textit{Chisel} underscores the defensive imperative to augment perimeter controls with behavioral analytics that can detect protocol tunneling in real time without relying on static signatures alone.

\subsubsection{LoLBins}\label{lolbins}
\textbf{\textit{Living-off-the-land binaries and scripts}} (\textbf{\textit{LOLBins}}/\textbf{\textit{LOLBAS}}) are legitimate executables that ship with, or are routinely installed on, modern operating systems yet can be subverted to perform malicious tasks\footnote{https://www.connectwise.com/blog/unveiling-lolbins-living-off-the-land-binaries}. Because no foreign code is introduced, their invocation blends into normal administrative activity and eludes traditional application-whitelisting or signature-based controls, a property that has made them central to contemporary \textit{living-off-the-land} tradecraft. Industry reporting over the past two years consistently ranks \textit{PowerShell}, \textit{cmd.exe}, \textit{rundll32.exe} and \textit{PsExec} among the most frequently abused specimens, underscoring the defensive challenge posed by binaries that are indispensable to system administration while simultaneously weaponizable for \textit{Post-Exploitation} objectives. 

\begin{figure}[H]
    \centering
    \frame{\includegraphics[width=0.8\linewidth]{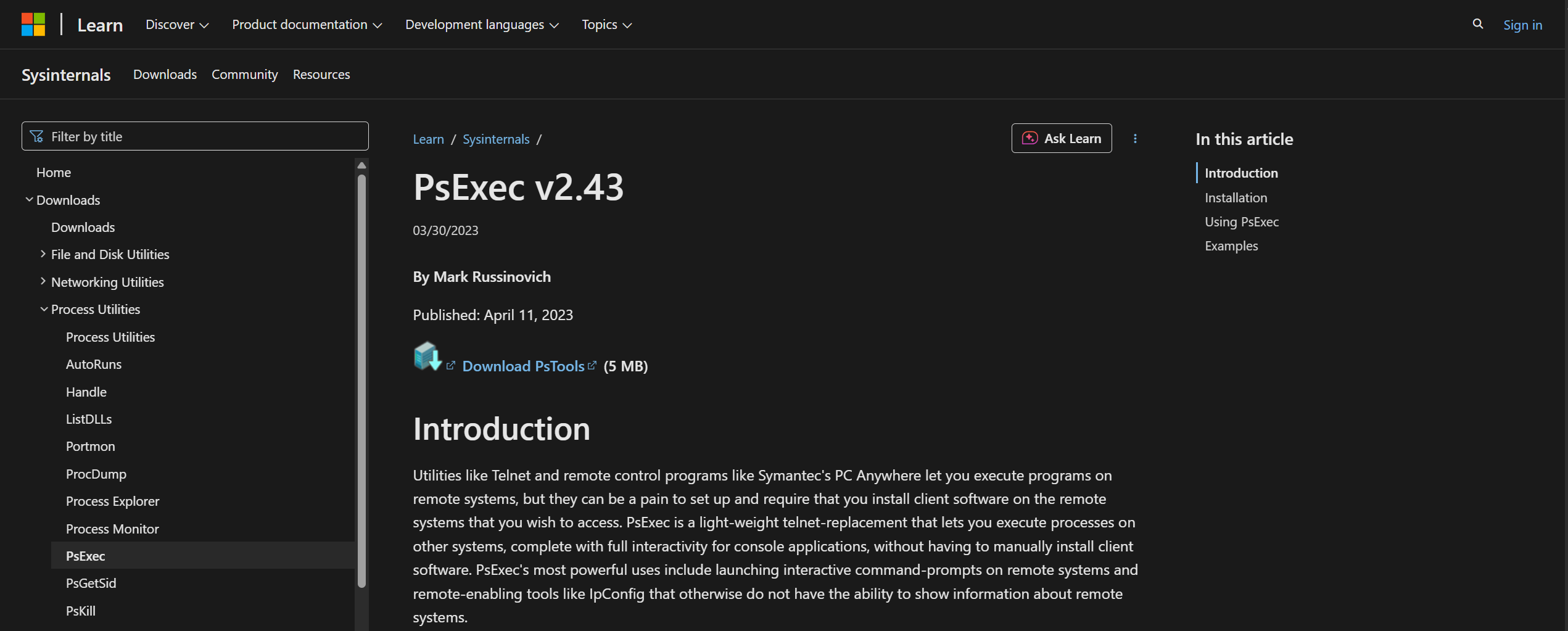}}
    \caption{PsExec's \href{https://learn.microsoft.com/en-us/sysinternals/downloads/psexec}{Microsoft Learn} Page.}
    \label{fig:3}
\end{figure}

\textbf{\textit{PsExec}}, released by \textit{Sysinternals} and now maintained under \textit{Microsoft Learn} (Fig. \ref{fig:3}), epitomizes this dual-use tension. In legitimate workflows it provides a lightweight, interactive alternative to \textit{Telnet} or \textit{RDP}, enabling administrators to execute commands on remote hosts without a resident agent; its single executable can be copied and invoked ad hoc, and it inherits the target user’s standard I/O for seamless console redirection. 

Operationally, the tool achieves remote execution by authenticating over SMB, pushing a helper binary named \textit{PSEXESVC.exe} into the target’s \textit{ADMIN\$} share, registering that helper as a temporary Windows service, and marshaling all subsequent input and output through the named pipe \textit{\textbackslash\textbackslash.\textbackslash pipe\textbackslash psexesvc}\footnote{https://redcanary.com/blog/threat-detection/threat-hunting-psexec-lateral-movement/}. Even if the original file is renamed, these characteristic artifacts remain constant: the service image path, the pipe identifier, the internal file metadata strings \textit{PsExec} and \textit{PsExec Service Host}, and an optional registry key under \textit{HKCU\textbackslash Software\textbackslash Sysinternals\textbackslash PsExec} that records EULA acceptance.

The binary’s prevalence in \textit{Threat Reporting} is substantial. The joint \textit{CISA–FBI} advisory on \textit{Play ransomware}, last updated 4 June 2025, lists \textit{PsExec} among the principal tools the group uses for lateral movement and on-disk encryption orchestration\footnote{https://www.cisa.gov/news-events/cybersecurity-advisories/aa23-352a}. \textit{Symantec}’s June 2025 analysis of a \textit{Fog ransomware intrusion} likewise documents \textit{PsExec} being invoked with the \textit{-h -s} flags to delete forensic artifacts and to relay secondary backdoors, demonstrating how its signed status and administrative lineage help attackers both traverse and sanitize victim environments\footnote{https://www.security.com/threat-intelligence/fog-ransomware-attack}. 

\textit{PsExec}’s ecosystem now includes open-source or proprietary clones such as \textit{RemCom}, \textit{PAExec} and \textit{CSExec}, each replicating the same S\textit{MB-push-service} pattern but substituting bespoke helper binaries and pipe names. Because these derivatives inherit the underlying communication model, defenders can generalize detection by focusing on anomalous service creation events coupled with short-lived executables dropped into \textit{ADMIN\$} or by enumerating named pipes that deviate from an established baseline of domain services. Studies by \textit{Red Canary} and \textit{DFIR} practitioners further note that even obfuscated builds must still negotiate SMB session-setup traffic, offering an additional network-based vantage point for hunting. 

\textit{PsExec} exemplifies the broader \textit{LOLBin} dilemma: a single, signed, and widely trusted executable whose legitimate utility is matched only by its utility to adversaries. Its deterministic service-based workflow and indelible on-disk and in-memory artifacts furnish defenders with deterministic signals, yet those signals will only surface when host, network and registry instrumentation are jointly interrogated—a reminder that \textit{living-off-the-land} threats are best countered with \textit{living-off-the-land} telemetry.

\subsubsection{Doppelganger and RTCore64.sys}\label{doppel}
\begin{figure}[H]
    \centering
    \frame{\includegraphics[width=0.7\linewidth]{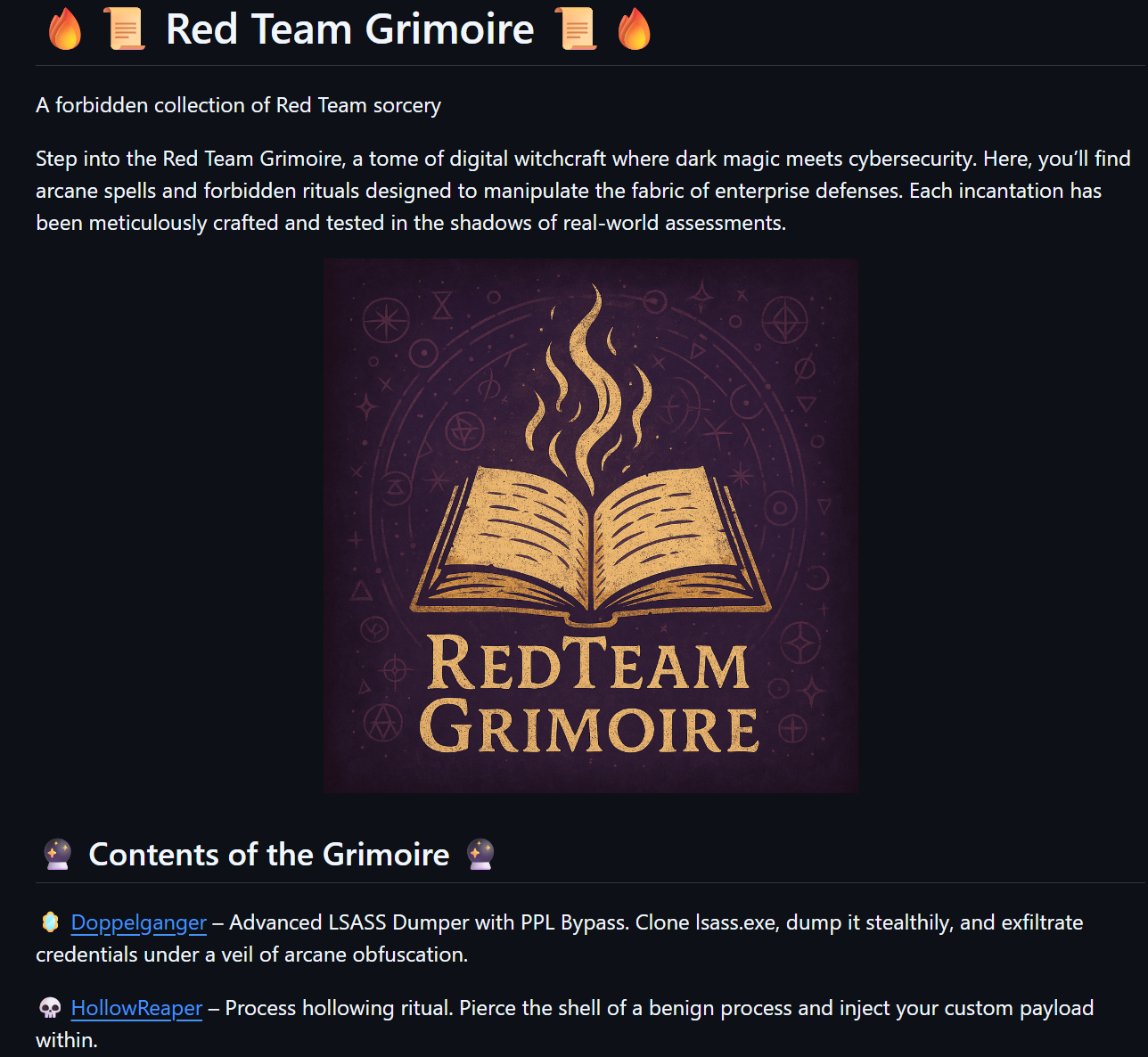}}
    \caption{Doppelganger public \href{https://github.com/vari-sh/RedTeamGrimoire}{GitHub Repository}.}
    \label{fig:4}
\end{figure}

\textbf{\textit{Doppelganger}} is a contemporary credential-access utility released in early 2025 by the researcher \textit{vari.sh}\footnote{https://github.com/vari-sh/RedTeamGrimoire/tree/main/Doppelganger} (Fig. \ref{fig:4}). Its declared objective is to extract the contents of the \textit{Local Security Authority Subsystem Service} on Windows 10 and 11 even when the process is shielded by \textit{Protected Process Light} (\textit{PPL}), \textit{Virtualisation-Based Security} and modern \textit{Endpoint Detection and Response} hooks. Unlike classic dumpers that attach to \textit{LSASS} and invoke \textit{MiniDumpWriteDump} directly, \textit{Doppelganger} first forges a near-perfect clone of the live \textit{LSASS} process with the undocumented system call \textit{NtCreateProcessEx}; the clone inherits the entire address-space of its parent, including logon sessions, ticket caches and \textit{NTLM} hashes, yet it does not inherit the \textit{PPL} flag, nor the user-mode callbacks planted by most \textit{EDR} products. All subsequent memory operations therefore occur in an unprotected context and elude the heuristics that watch for handles opened against the real LSASS.

The obstacle that remains is kernel enforcement of \textit{PPL} on the original process, because Windows will refuse to create a full clone of a protected target. \textit{Doppelganger} removes that obstacle by importing a vulnerable, but legitimately signed, kernel driver— \textbf{\textit{RTCore64.sys}}, the hardware-monitoring component shipped with \textit{MSI Afterburner}. That driver is affected by \textbf{CVE-2019-16098}, an \textit{arbitrary read-write flaw} in its \textit{IOCTL} implementation that allows any authenticated user to \textit{read or overwrite kernel memory}. Because the binary is \textit{Microsoft-signed}, \textit{Windows} loads it without complaint; once resident, its device object exposes unrestricted primitives that let user-mode callers modify executive data structures from ring 3\footnote{https://www.picussecurity.com/resource/blog/blackbyte-ransomware-bypasses-edr-products-via-rtcore64.sys-abuse}.

\textit{Doppelganger} leverages those primitives to traverse the doubly linked list of \textit{EPROCESS} structures, locate the node corresponding to \textit{lsass.exe}, and patch the three contiguous bytes that encode the process-protection level. By zeroing \textit{SignatureLevel}, \textit{SectionSignatureLevel} and \textit{Protection} it momentarily strips \textit{LSASS} of \textit{PPL} status. The tool then invokes \textit{NtCreateProcessEx} to fork the unprotected address-space and immediately restores the original protection bytes, ensuring that forensic artifacts are limited to a brief twelve-byte window of altered kernel memory. All reads and writes are funneled through \textit{RTCore64}’s \textit{DeviceIoControl} interface, so no user-mode \textit{API} that defenders typically hook—\textit{OpenProcess}, \textit{ReadProcessMemory}, \textit{NtWriteVirtualMemory}—is ever called. This tactic mirrors the broader \textbf{Bring-Your-Own-Vulnerable-Driver} (\textbf{BYOVD}) trend observed in \textit{BlackByte} and other ransomware families that likewise employ \textit{RTCore64.sys} to blind \textit{EDR} kernel callbacks.

Once the clone exists, \textit{Doppelganger} acquires a duplicated \textit{SYSTEM token} from a trusted process such as \textit{winlogon.exe}, enables \textit{SeDebugPrivilege}, and invokes a customized build of \textit{MiniDumpWriteDump} against the clone. \textit{I/O callbacks} redirect every chunk into a heap buffer, so nothing is written to disk in clear text; the resulting blob is \textit{XOR}-obfuscated in memory and only the cipher-text dump is persisted, frustrating pattern-based file scanners. A companion \textit{Python script} in the project decrypts the dump offline for analysis. Throughout the operation, Windows telemetry records only a transient service load for \textit{RTCore64.sys} (unless service creation is further obfuscated), a handful of \textit{DeviceIoControl} calls to an otherwise legitimate driver, and the short-lived presence of an unsigned executable named \textit{Doppelganger.exe}.

The \textit{Doppelganger}/\textit{RTCore64} pairing therefore illustrates the state of the art in post-exploitation credential theft: it embraces a signed, high-integrity driver to gain kernel primitives, applies surgical one-bit modifications that are reverted within milliseconds, and moves all heavy processing to an unprotected surrogate so that the real \textit{LSASS} remains inviolate and fully monitored—to no avail. Effective counter-strategy depends on \textit{driver-blocklisting} with \textit{HVCI} or Microsoft’s vulnerable-driver list, continuous auditing of service deployments, and kernel-level sensors able to detect protection-field toggling rather than classical user-mode dump behavior.

\subsubsection{Targeting Active Directory}\label{ad}
\textbf{\textit{DumpAADUserRPT}} is a compact \textit{C-sharp} utility that re-implements the \textit{Get-AADInt UserPRTToken} routine from the \textit{AADInternals} project in order to exfiltrate the browser-scoped cookie \textit{x-ms-RefreshTokenCredential}, better known as the \textit{Azure Active Directory Primary Refresh Token} (\textit{PRT}), from Windows hosts that are \textit{Azure AD–joined} or \textit{hybrid-joined}\footnote{https://github.com/Hagrid29/DumpAADUserRPT}. Because a \textit{PRT} can be exchanged for fresh \textit{OAuth2} access tokens for any application registered in the tenant, possession of a valid \textit{PRT} is functionally equivalent to long-term interactive control over the user’s cloud identity; Microsoft likens it to a device-bound \textit{Kerberos TGT}, but in practice it often bypasses conditional-access prompts and even multi-factor requirements when presented from the same device context.

The tool first obtains a one-time nonce—mandatory since Microsoft’s October 2020 hardening—by invoking its \textit{get\_nonce mode}, then uses that nonce to request the \textit{PRT}. It offers three acquisition paths, each mirroring a legitimate \textit{SSO} workflow. The most direct path spawns \textit{BrowserCore.exe}, supplies a \textit{JSON} blob containing the nonce on standard input, and receives the \textit{PRT} on standard output; \textit{BrowserCore} mediates the request through the undocumented COM interface exposed by \textit{MicrosoftAccountTokenProvider.dll}. \textit{Microsoft Defender for Endpoint} now flags this straight-through invocation, so \textit{DumpAAD\newline UserRPT}’s second path emulates the \textit{Windows 10 Accounts Chrome} extension: it fabricates the two named pipes that Chrome would normally create, launches \textit{BrowserCore} with its characteristic \textit{--parent-window} command line, and shuttles the request through those pipes, a sequence that defeats the stock \textit{MDE} heuristic because the telemetry is indistinguishable from genuine browser activity. The third path dispenses with \textit{BrowserCore} altogether and drives the \textit{COM} provider directly—an approach first documented by Lee Christensen—which yields the same cookie with fewer forensic artifacts and no dependence on named pipes\footnote{https://bloodhoundenterprise.io/blog/2020/07/14/requesting-azure-ad-request-tokens-on-azure-ad-joined-machines-for-browser-sso/}.

Once the \textit{Base64-encoded PRT} is in hand the attacker may set it as the value of \textit{x-ms-RefreshTokenCredential} in any browser session directed to \textit{login.microsoftonline.com}, thereby inheriting the victim’s \textit{single-sign-on} posture; alternatively, the token can be fed back into \textit{AADInternals} to mint arbitrary \textit{Graph} or \textit{MSAL} access tokens without further contact with the compromised endpoint. Because the \textit{PRT} embeds both \textit{device ID} and user \textit{SID}, it remains valid until the device object is disabled or the user’s refresh-token family is revoked, so it is routinely harvested for lateral movement in cloud-heavy environments.

\textbf{\textit{Whisker}} is a compact C-sharp tool released by \textit{Elad Shamir} in 2021 that operationalization the \textit{shadow-credentials} technique, a takeover primitive that abuses the \textit{Active Directory} attribute \textit{msDS-KeyCredentialLink}\footnote{https://github.com/eladshamir/Whisker}. By appending a rogue \textit{KeyCredential} object to that multi-value attribute on a user or computer account, an operator secretly endows the target with an additional public-key identity and can thereafter authenticate as that object with \textit{Kerberos PKINIT}, all without resetting a password, disabling pre-authentication, or touching service-principal names. Because the injected key material is treated by the domain controller exactly like the trusted keys produced during \textit{Windows Hello for Business} provisioning, the compromise persists through password changes and survives most incident-response playbooks that focus on credential hygiene. \textit{Whisker} draws its low-level directory-manipulation routines from \textit{Michael Grafnetter}’s \textit{DSInternals} library and surfaces them in a single-binary \textit{CLI} that supports adding, listing, deleting or wholesale clearing of \textit{msDS-KeyCredentialLink} values.

The attack relies on infrastructure introduced with \textit{Windows Server 2016}. Under the \textit{Key-Trust} variant of \textit{Windows Hello}, a client’s \textit{TPM-backed} public key is stored in \textit{msDS-KeyCredentialLink} and later used for \textit{public-key pre-authentication}. If an adversary can write to that attribute, the domain controller will accept the rogue key and issue \textit{Ticket-Granting Tickets} signed for the target principal, provided the \textit{DC} itself holds a server-authentication certificate—normally auto-enrolled in any forest that runs \textit{AD CS}. \textit{Whisker} programmatically generates a \textit{self-signed certificate}, packages the public key together with a randomly generated \textit{DeviceID GUID }and ancillary metadata, \textit{ASN.1}-encodes the structure into a \textit{KeyCredential} blob, and commits the modification over \textit{LDAP}. The tool then prints a ready-made \textit{Rubeus} command that requests a \textit{TGT} with the newly minted key pair, proving immediate control\footnote{https://posts.specterops.io/shadow-credentials-abusing-key-trust-account-mapping-for-takeover-8ee1a53566ab}.

Practical deployment demands only two pre-conditions: at least one \textit{Server 2016} or newer domain controller equipped for \textit{PKINIT}, and an access path that grants the attacker \textit{WRITE\_PROPERTY} permission on the victim object or a higher-level right such as \textit{GenericWrite} or \textit{GenericAll} obtained through mis-delegation, \textit{SID-history}, or compromise of a sufficiently privileged principal. No \textit{domain-wide misconfiguration} is necessary; a single computer account that inherits an overly permissive \textit{Access Control Entry} is enough to elevate to \textit{Domain Admin} in many \textit{real-world forests}. Laboratory walk-throughs show low-privilege users adding a shadow credential to a domain-admin computer account and moments later mounting the \textit{DC}’s \textit{C\$} share under the computer’s identity, illustrating both lateral movement and privilege escalation in one step.

The forensic footprint is proportionally small. On the wire, the \textit{LDAP Modify request} differs from routine management traffic only in that it targets \textit{msDS-KeyCredentialLink}, a field rarely set outside \textit{Windows Hello enrollment}. On the host, no service restarts occur, and the original password and password-history remain intact, so traditional indicators such as Event 4741/4723 are absent. Once the rogue key is in place, \textit{Kerberos tickets} issued to it are indistinguishable from legitimate \textit{PKINIT} tickets unless the \textit{KDC} audit log is configured to surface the \textit{KdcProxyCertHash} field. \textit{Elastic}, \textit{Microsoft} and \textit{independent} researchers therefore recommend \textit{DS} Access \textit{auditing—Event 4662} with property \textit{GUID 6f9675b5-8deb-4f10-9d5d-6052f4dba6af}—combined with periodic directory sweeps that flag objects whose \textit{msDS-KeyCredentialLink} was modified outside normal \textit{Hello enrollment endpoints}.

In sum, Whisker translates an esoteric attribute abuse into a turnkey post-exploitation primitive that confers stealthy, long-lived control over any \textit{Active Directory} object whose \textit{ACL} can be modified. The technique exemplifies a broader shift toward key-material attacks that bypass password semantics altogether and reinforces the need for fine-grained directory permission hygiene and comprehensive change auditing in modern enterprise forests.

\newpage
\subsection{IoCs, TTPs \& Yara Rules} \label{App:IoC}
The entire set of \textit{IoCs}, \textit{TTPs} and few \textit{Yara} Rules, gathered through-out this entire analysis, are available inside the following \textit{AlienVault OTX} \href{https://otx.alienvault.com/pulse/6862d9f230c3c621c39d1617}{pulse}.

\begin{figure}[H]
    \centering
    \frame{\includegraphics[width=1\linewidth,frame]{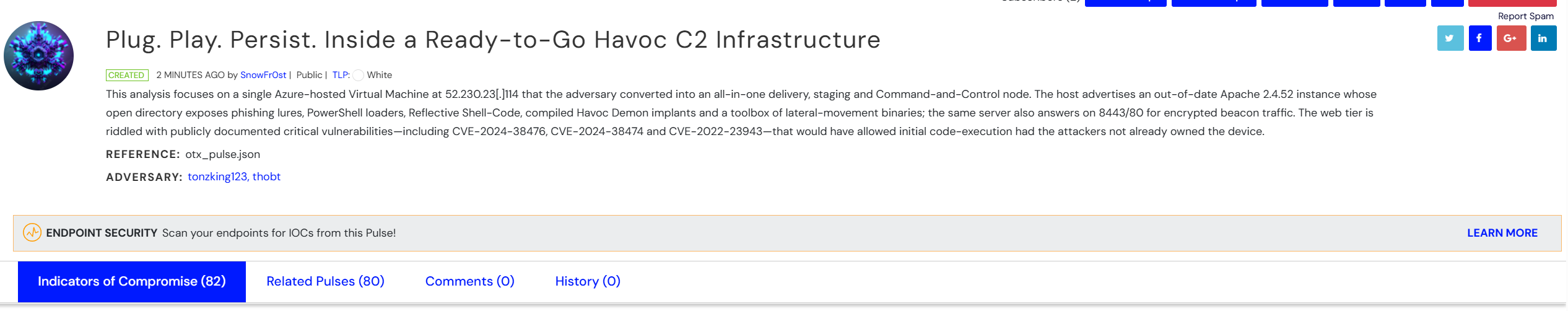}}
    \caption{Overview of the \textit{AlienVault OTX pulse}}
    \label{fig:165}
\end{figure}

\newpage
\subsection{Sigma Rules}
\vspace*{\fill} % Push content to the center vertically
\begin{center}
\begin{lstlisting}[language=yaml]
title: Havoc Demon - DEADBEEF Framing Constant
id: 9ff2fe49-4b53-4186-8f6e-3d9e0e1eab2a
status: experimental
description: Detects outbound Demon beacons that start with the 0xDEADBEEF marker often followed by AgentID and length fields.
references:
  - https://github.com/HavocFramework/Havoc
author: Alessio Di Santo
logsource:
  product: network
  category: proxy
detection:
  selector:
    DestinationPort: 8443
    DestinationIp|endswith: ".114"      # 52.230.23.114 in the paper
    PayloadHex|startswith: "DEADBEEF"
  condition: selector
falsepositives:
  - Very rare; some network-debug tools use the same sentinel bytes.
level: high
tags:
  - attack.c2
  - framework.havoc
\end{lstlisting}
\end{center}
\vspace*{\fill} % Push content to the center vertically
\newpage
\vspace*{\fill} % Push content to the center vertically
\begin{center}
\begin{lstlisting} [language=yaml]
title: IFEO Persistence - Havoc / Meterpreter-style Loader
id: c4d4b3e3-b808-4e53-9835-52f9b97df2c3
status: stable
description: Flags creation or modification of IFEO Debugger values, a technique used by the loader outlined in the report to launch reflective shell-code on application start-up.
author: Alessio Di Santo
logsource:
  product: windows
  service: security
  definition: EventID 4657 or 13 (Sysmon Registry)
detection:
  selector_4657:
    EventID: 4657
    ObjectName|contains: '\\Image File Execution Options\\'
    Details|contains: 'Debugger'
  selector_sysmon:
    EventID: 13
    TargetObject|contains: '\\Image File Execution Options\\'
    Details|contains: 'Debugger'
  condition: selector_4657 or selector_sysmon
falsepositives:
  - Legitimate debuggers or application compatibility shims during software development.
level: medium
tags:
  - attack.persistence
  - attack.t1546.012      # IFEO Injection
\end{lstlisting}
\end{center}
\vspace*{\fill} % Push content to the center vertically
\newpage
\vspace*{\fill} % Push content to the center vertically
\begin{center}
\begin{lstlisting}[language=yaml]
title: Active Directory Shadow Credentials (Whisker / Key Trust Abuse)
id: 0b5d0de5-baf5-4dad-9e5e-5efc93db7b7e
status: stable
description: Detects modifications to msDS-KeyCredentialLink, typically exploited by Whisker to create hidden device credentials that grant Domain-Admin equivalent access.
author: Alessio Di Santo
logsource:
  product: windows
  service: security
detection:
  selector:
    EventID: 4662
    ObjectType: '6f9675b5-8deb-4f10-9d5d-6052f4dba6af'   # msDS-KeyCredentialLink
    Properties|contains: 'msDS-KeyCredentialLink'
  condition: selector
falsepositives:
  - Legitimate Windows Hello for Business enrolment.
level: high
tags:
  - attack.privilege_escalation
  - attack.t1550.003      # AD CS / certificate abuse
\end{lstlisting}
\end{center}
\vspace*{\fill} % Push content to the center vertically
\newpage
\subsection{Infection Chain}
\begin{figure}[H]
    \centering
    \frame{\includegraphics[width=0.9\textheight, angle=270,frame]{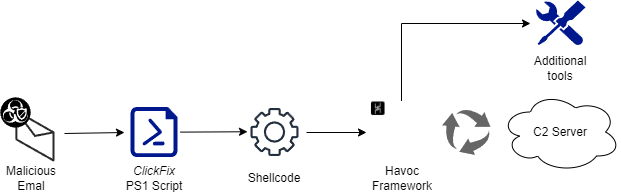}}
    %\caption{This is a rotated image using `graphicx`.}
\end{figure}

\subsection{Diamond Model}
\begin{figure}[H]
    \centering
    \frame{\includegraphics[width=0.95\textheight, angle=270,frame]{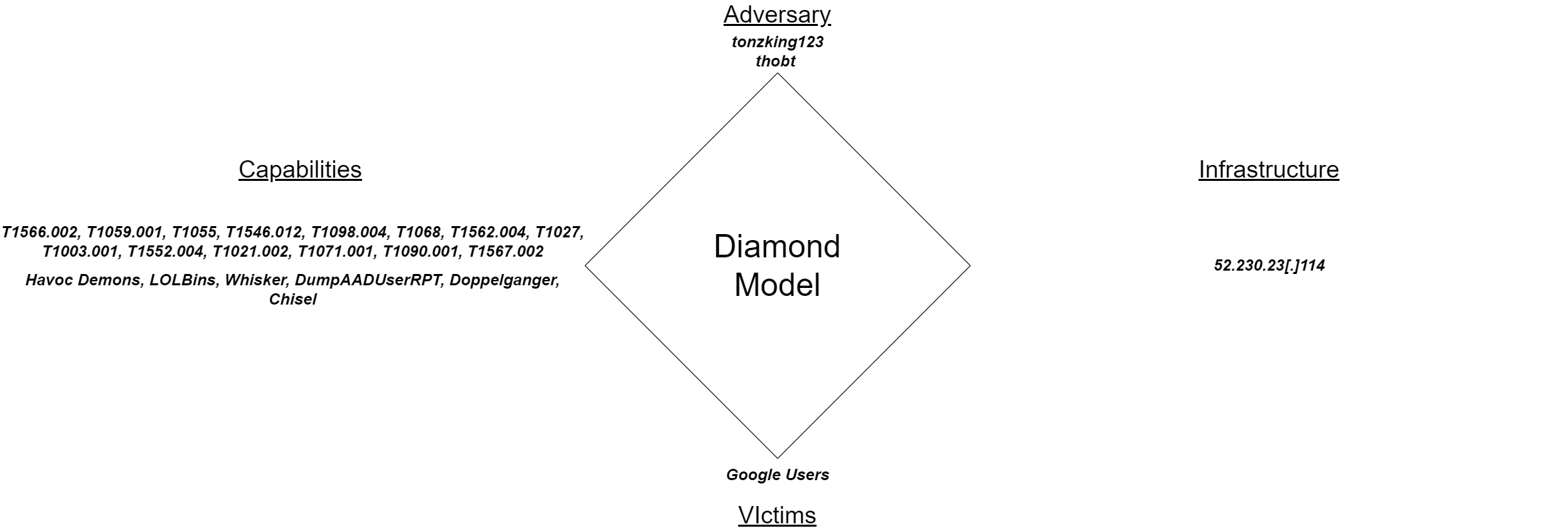}}
    \label{fig:DM}
\end{figure}

\end{document}